\documentclass[aps,pra,twocolumn,floatfix,superscriptaddress]{revtex4-1}

\usepackage{mathrsfs}
\usepackage{graphicx}
\usepackage[english]{babel}
\usepackage{amsmath}
\usepackage{amssymb}

\usepackage{relsize}
\usepackage{CJK}
\usepackage{dsfont}
\usepackage{bm}

\newcommand{\me}{\mathrm{e}}
\newcommand{\mi}{\mathrm{i}}

\newcommand{\dif}{\mathrm{d}}

\allowdisplaybreaks
\begin{document}

\title{Finite-temperature topological phase transitions of spin-$j$ systems in Uhlmann processes: General formalism and experimental protocols}
\author{Xu-Yang Hou }
\affiliation{Department of Physics, Southeast University, Jiulonghu Campus, Nanjing 211189, China}
\author{Hao Guo }
\email{guohao.ph@seu.edu.cn}
\affiliation{Department of Physics, Southeast University, Jiulonghu Campus, Nanjing 211189, China}
\author{Chih-Chun Chien}
\email{cchien5@ucmerced.edu}
\affiliation{Department of physics, University of California, Merced, CA 95343, USA}

\begin{abstract}
The Uhlmann process is built on the density matrix of a mixed quantum state and offers a way to characterize topological properties at finite temperatures. We analyze an ideal spin-$j$ quantum paramagnet in a magnetic field undergoing an Uhlmann process and derive general formulae of the Uhlmann phase and Loschmidt amplitude for arbitrary $j$ as the system traverses a great circle in the parameter space.
A quantized jump of the Uhlmann phase signifies a topological quantum phase transition (TQPT) of the underlying process, which is accompanied by a zero of the Loschmidt amplitude. The exact results of $j=\frac{1}{2}$ and $j=1$ systems show topological regimes that only survive at finite temperatures but not at zero temperature, and the number of TQPTs is associated with the winding number in the parameter space. Our results pave the way for future studies on finite-temperature topological properties, and possible experimental protocols and implications for atomic simulators and digital simulations are discussed.
\end{abstract}

\maketitle

\section{Introduction}
Topological quantum systems, including topological insulators and topological superconductors~\cite{TKNN,Haldane,KaneRMP,ZhangSCRMP,MooreN,MoorePRB,FuLPRL,Bernevigbook,ChiuRMP,KaneMele,KaneMele2,BernevigPRL}, have become a major topic in condensed matter and atomic, molecular, and optical physics during the past few decades. The pioneer studies~\cite{Berry84,Bohm03,Vanderbilt_book,Cohen19} have been focused on the ground state, where the Berry phase and its curvature play an important role in characterizing the underlying topological properties. However, mixed quantum states are common in finite-temperature or out-of-equilibrium systems. Thus, it is natural to extent the study of the Berry phase (or other geometrical quantities) from pure states to mixed states.
Several attempts to reconcile the description of topological properties of mixed states have been proposed~\cite{Uhlmann86,GPMQS2,GPMQS1,HZUPPRL14,ViyuelaPRL14,2DMat15,Kempkes16,MeraPRL17,DiehlPRX17,UPPRB18}.
Among these candidates, the Uhlmann phase originally developed in Refs.~\cite{Uhlmann86,Uhlmann89,Uhlmann91,Uhlmann96} is a promising generalization of the Berry phase to finite-temperature systems. Similar to the Berry phase, the Uhlmann phase is a geometrical phase produced when a mixed quantum state is `parallel-transported' during a cyclic process, which will be referred to as the Uhlmann process in the following discussion. The original framework was built from a mathematical point of view~\cite{Uhlmann86}, causing some difficulties for a deep understanding of the physical implications of the Uhlmann phase in the physics community after its debut.

Recently, the Uhlmann phase has drawn attention as the research of topological systems moves beyond the ground-state formalism. Early applications of the Uhlmann phase to topological systems have showed promising topological phase transitions at finite temperatures~\cite{ViyuelaPRL14,HZUPPRL14,2DMat15}. Despite its resemblance of the formalism of the Berry phase, the Uhlmann phase has some distinct behavior. For example, the fiber bundle behind the Uhlmann phase is topologically trivial~\cite{DiehlPRB15} with vanishing topological characteristics. Therefore, the classification of different topological phases in the ground state according to the Chern number or other characteristics cannot be generalized to finite temperatures based on the Uhlmann bundle. Moreover, it is found that the Uhlmann process is incompatible with the dynamical process while the Berry process is compatible with the dynamical process~\cite{ourPRB20}. The incompatibility seems common in defining a geometric phase of mixed quantum states. For example, in the introduction of the interferometric geometric phase inspired by the Mach-Zehnder interferometry~\cite{GPMQS1}, the dynamical evolutionary condition is also excluded.

Despite the drawback from the fiber bundle, the Uhlmann phase remains useful in identifying the topological quantum phase transitions (TQPTs) of mixed quantum states because a jump of the Uhlmann phase at a point with vanishing overlap of the purified states indicates a change of topological properties~\cite{ourPRB20b}. Interestingly, the studies of dynamical quantum phase transitions (DQPTs) of mixed quantum states in a quench process may also be described by the overlap of purified states ~\cite{DQPT13,DQPT14,DQPT15,DQPTreview18}. It was found~\cite{ourPRB20b} that the TQPTs and DQPTs of mixed quantum states can be identified by the zeros of the Loschmidt amplitude from the overlap of the corresponding states. In this context, the argument of the Loschmidt amplitude respectively gives the Uhlmann phase and dynamical phase in the corresponding process. However, the dynamical phase does not carry topological information~\cite{ourPRB20}, so we will solely focus on the Uhlmann process here.

So far, our knowledge of the Uhlmann phase is less than that of the Berry phase, as only a handful of explicit expressions of the Uhlmann phase of certain systems are found. However, the Uhlmann phase has become a measurable quantity for simple two-level systems due to a protocol and its simulation on IBM's quantum computer~\cite{npj18}. A recent study derived the Uhlmann phase of a spin-$j$ paramagnet when an external magnetic field traverses the equator of the parameter space~\cite{Galindo21}, showing quantized jumps of the Uhlmann phase. Here we will derive the general expressions of the Uhlmann connection and Uhlmann fidelity, equivalent to the Loschmidt amplitude~\cite{DQPTreview18}, using a different framework and formalism. An analytic expression of the Uhlmann phase of the spin-$j$ system when the system traverses a circle of longitude or the equator of the parameter space multiple times will be shown. We present concrete examples of $j=1/2$ and $j=1$ systems, showing finite-temperature topological quantum phase transitions (TQPTs) with vanishing Loschmidt amplitude and quantized jumps of the Uhlmann phase. Moreover, the examples show that the number of TQPT points is related to the winding number in the parameter space. An interesting phenomenon of the spin-$j$ system is the emergence of an intermediate-temperature topological phase sandwiched by topologically trivial phases at both high and low temperatures. Such a phenomenon offers an example that temperature may not always destroy topological properties. Since the spin-$j$ system may be realized in quantum simulators or tested on quantum computers, we will present feasible experimental protocols and discuss implications of the TQPTs from an Uhlmann process.

The rest of this paper is organized as follows. In Sec.\ref{sec.2}, we briefly review the theoretical foundation of the Uhlmann phase and the spin-$j$ systems. In Sec.\ref{sec.3}, we derive the general expression of the Uhlmann connection for spin-$j$ systems. In Sec.\ref{sec.4}, we further derive the formulae of the Uhlmann fidelity/Loschmidt amplitude for Uhlmann processes along the circles of longitude and the equator. In Sec.\ref{sec.5}, we apply our result to explicit examples with $j=\frac{1}{2}$ and 1 systems and present the numerical analysis. Finally, in Sec.\ref{sec.6}, we briefly discuss possible experimental measurement protocols for measuring the Uhlmann phase using quantum simulators or quantum computers. Some details and derivations are summarized in the Appendix.

\section{Theoretical background}\label{sec.2}
\subsection{Uhlmann process}
In the following, we set $\hbar=1$ and $k_B=1$ and briefly review the purification of a density matrix. Following Uhlmann's approach~\cite{Uhlmann86,Uhlmann89}, a full-rank density matrix $\rho$ can be factorized as
\begin{align}\label{Wr}
\rho=WW^\dagger,
\end{align}
where $W$ is called the purification or amplitude of $\rho$. Conversely, a given full-rank matrix $W$ can be uniquely decomposed as $W=\sqrt{\rho}U$ where the unitary matrix $U$ is analogous to the phase factor of a wavefunction. The amplitudes form a Hilbert space $H_W$ endowed with an inner product known as the Hilbert-Schmidt product: $(W_1,W_2)\equiv \text{Tr}(W^\dagger_1W_2)$. In general, the inner product applies to any two purifications of the same dimension. However, we will mainly consider $W_1$ and $W_2$ as purifications of the same density matrix in the following. The purification $W$ can be constructed by the eigenvectors $|i\rangle$ of $\rho$ from the Hilbert space $\mathcal{H}$, where the collection $\{|i\rangle\}$ serves as a basis. However, it can be cast into the form of a pure state $|W\rangle$ by introducing a tensor-product space. Explicitly, there is a one-to-one mapping between the outer-product expression and the tensor-product expression:
\begin{align}\label{w2}
W=\sum_i\sqrt{\lambda_i}|i\rangle\langle i|U\leftrightarrow|W\rangle=\sum_i\sqrt{\lambda_i}|i\rangle\otimes U^\text{T}|i\rangle,
\end{align}
where $\lambda_i$ is the $i$-th eigenvalue of $\rho$, and the superscript ``T'' denotes the transpose with respect to the eigenbasis of $\rho$. $|W\rangle$ is called a purified state of $\rho$ and spans the Hilbert space $\mathcal{H}\otimes\mathcal{H}$, where the second $\mathcal{H}$ is a copy of the first one and is referred to as the ancillary system.
According to Eq.~(\ref{w2}), it can be shown that the conventional inner product between two purified states in $\mathcal{H}\otimes\mathcal{H}$ reproduces the the Hilbert-Schmidt product between two purifications in $H_W$: $\langle W_1|W_2\rangle=(W_1,W_2)$. Moreover, the density matrix can be recovered by tracing out the ancilla degrees of freedom: $\rho=\textrm{Tr}_2(|W\rangle\langle W|)$.
In such a way, a mixed state is equivalently represented by a pure state that is an entangled state from two subspaces consisting of the original system and the ancilla.
While the amplitude makes the theoretical derivations more straightforward, the purified state allows experimental simulations and probes of mixed states with the help of an ancilla. Both advantages will be demonstrated in this work.

When a system is controlled by some external parameters $\mathbf{R}\equiv(R_1,R_2,\cdots,R_k)$ spanning a parameter space $M$, the Hamiltonian and density matrix depend on those parameters. As the parameters traverse a curve $\gamma$ in the parameter space, the system varies accordingly.
We assume the curve $\gamma$ is parameterized by $t$, which is not necessary the time. According to Refs.~\cite{Uhlmann86}, an amplitude $W$ is said to be parallel-transported if it satisfies
\begin{align}\label{Wpt}
\dot{W}^\dagger W=W^\dagger\dot{W},
\end{align}
where $\dot{W}=\frac{\dif W}{\dif t}$. The condition preserves locally the parallelity between $W$ and its adjacent amplitude $W+\dif W$. Here the parallel condition between two amplitudes $W_{1,2}$ is given by \begin{align}\label{Wpt1}W^\dagger_1W_2=W^\dagger_2W_1>0,\end{align} where `$>0$' means all the eigenvalues of the corresponding matrix are positive.
Note Eq.~(\ref{Wpt}) is the differential form of the condition (\ref{Wpt1}).
The condition further leads to $(W_1,W_2)=(W_2,W_1)>0$, and the inner product is called the fidelity, which measures the similarity between $W_1$ and $W_2$. Under the parallel-transport condition, the Hilbert-Schmidt distance between $W_1$ and $W_2$ is minimized~\cite{Uhlmann86,Uhlmann89}.

The Uhlmann process is a cyclic process, where an amplitude is parallel-transported along a closed path in the parameter space. Here `cyclic' means the initial and final density matrices are the same. Although the parallelity is maintained during an Uhlmann process, the final amplitude $W_\text{f}$ may not be the same as the initial amplitude $W_\text{i}$ since the parallel condition lacks transitivity, even though $W_\text{i}$ and $W_\text{f}$ are two purifications of the same density matrix.
For a cyclic Uhlmann process, the overlap $(W_\text{i},W_\text{f})$ is called the Uhlmann fidelity and may be a complex number. Let $W_\text{i}=W(0)$ and $W_\text{f}=W(\tau)$, where $\tau$ is the length of curve in the parameter space parametrized by $t$. The Uhlmann process may also be described by using the concept of fiber bundle, and Sec.~\ref{app:fb} gives a brief overview.

\subsection{Uhlmann fidelity and Uhlmann phase}
The Uhlmann process requires the density matrix of a system to remain in equilibrium through out the process. It has been shown that an Uhlmann process is not compatible with the dynamic process solely determined by the system Hamiltonian~\cite{ourPRB20}, implying the open-system nature of the Uhlmann process. When describing the evolution of the amplitude of the density matrix in an Uhlmann process, the system should be kept in equilibrium. This can be achieved by considering a system in contact with a reservoir and following quasi-stationary processes. The temperature $T$ is determined by the reservoir.
The Uhlmann fidelity is then given by~\cite{ourPRB20b}
\begin{align}\label{UF}
\mathcal{G}^U (T,\tau)=( W(0),W(\tau))=\text{Tr}[\rho(0)U(\tau)U^\dagger(0)].
\end{align}
The phase factor $U(\tau)$ is obtained by the parallel-transport condition
\begin{equation}
U(\tau)=\mathcal{P}\me^{-\oint_\tau A_U}U(0)
\end{equation}
where $A_U=-\dif UU^\dagger$ is the Uhlmann connection, and $\mathcal{P}$ denotes the path-order. Thus, the Uhlmann fidelity is
 \begin{align}\label{GrU0}
\mathcal{G}^U (T,\tau)=\text{Tr}\left[\rho(0)\mathcal{P}\me^{-\oint_\tau A_U}\right],
\end{align}
where $\rho(0)=\frac{1}{Z}\me^{-\beta H}$ with $\beta=\frac{1}{k_B T}$ and $Z=\text{Tr}(\me^{-\beta H})$ being the partition function.
Although $W(0)$ and $W(\tau)$ give the same density matrix $\rho(0)$, they differ by an Uhlmann holonomy element $\mathcal{P}\me^{-\oint_\tau A_U}$.
The Uhlmann connection can be cast in the form
given by~\cite{Uhlmann89,Uhlmann91}
\begin{align}\label{GrAU}
A_U=-\sum_{m,n}|\psi_m\rangle\frac{\langle \psi_m|[\mathrm{d}\sqrt{\rho},\sqrt{\rho}]|\psi_n\rangle}{\lambda_m+\lambda_n}\langle \psi_n|.
\end{align}
Here $\lambda_m$ and $|\psi_m\rangle$ are the $m$-th eigenvalue and eigenvector of $\rho$, respectively. The expression of the Uhlmann connection in the fiber-bundle language will be given in Sec.~\ref{app:fb}.

In Ref.~\cite{ourPRB20b}, the similarity and difference between the dynamical and Uhlmann processes have been discussed in details. Since $\mathcal{G}^U$ is the overlap between the initial and final purified states, it is also known as the Loschmidt amplitude.
The Uhlmann phase is given by the argument of $\mathcal{G}^U$:
 \begin{align}\label{Up}
\theta_U=\arg\mathcal{G}^U=\arg\text{Tr}\left[\rho(0)\mathcal{P}\me^{-\oint_\tau A_U}\right].
\end{align}
Since the phase $\theta_U$ is dimensionless, one can parametrize the loop in the parameter space by its length so that $\tau=1$.  Physically, this is equivalent to  adjusting the evolution rate of the Uhlmann process.

The Uhlmann phase is a generalization of the Berry phase to finite-temperature systems~\cite{ViyuelaPRL14,Asorey19}. As $T\rightarrow 0$, the weight factor of the ground state is infinitely larger than that of any excited state since $\beta\rightarrow\infty$. The Uhlmann process is then dominated by the cyclic process experienced by the ground state, a pure state. Thus, the Uhlmann phase approaches the Berry phase as $T\rightarrow 0$.

If the initial and final amplitudes lead to $\mathcal{G}^U_\rho=0$, the value of Uhlmann phase jumps, indicating the occurrence of a TQPT. Similar to the emergence of nonanalytic behavior in the dynamical free-energy across a DQPT induced by a quantum quench~\cite{DQPTreview18}, a TQPT can also be identified by the divergence of the geometrical generating function~\cite{ourPRB20b}
 \begin{align}\label{g}
g=-\lim_{L\rightarrow\infty}\frac{1}{L}\ln|\mathcal{G}^U|^2,
\end{align}
where $L$ is the degrees of freedom of the system. Ref.~\cite{ourPRB20b} offers a unified view of DQPTs and TQPTs through the vanishing of the Loschmidt amplitude. As explained in Refs.~\cite{DiehlPRB15,Asorey19}, the Uhlmann bundle may be viewed as a principle bundle with a global section, rendering it a trivial bundle with zero characteristics. A discussion of the Uhlmann curvature can be found in Appendix~\ref{app:curvature}.

\subsection{Spin-$j$ systems}
The Uhlmann process will be applied to a quantum spin-$j$ paramagnet in the presence of a magnetic field. Here we briefly review the system by considering an ensemble of spin-$j$ paramagnets influenced by an external magnetic field $\mathbf{B}$ with constant magnitude $B=|\mathbf{B}|$. The Hamiltonian is given by
\begin{align}\label{sjH}
H=\omega_0\hat{\mathbf{B}}\cdot \mathbf{J},
\end{align}
where $\omega_0$ is the Larmor frequency, $\hat{\mathbf{B}}=\mathbf{B}/B$, and $\mathbf{J}$ is the spin angular momentum. The controlled magnetic field $\mathbf{B}$ has a varying orientation characterized by the angles $\theta$ and $\phi$ via $\mathbf{B}=B(\sin\theta\cos\phi, \sin\theta\sin\phi,\cos\theta)^\text{T}$.
Thus, the parameter space $M$ in this case is the 2D sphere $S^2$.
A straightforward calculation shows that the Hamiltonian (\ref{sjH}) can be expressed in a parameter-dependent form,
\begin{align}\label{sjH1}
H(\theta,\phi)
&=\omega_0\me^{-\mi\phi J_z}\me^{-\mi\theta J_y}J_z\me^{\mi\theta J_y}\me^{\mi\phi J_z}\notag\\
&=R(\theta,\phi)\omega_0J_zR^\dagger(\theta,\phi)\notag\\
&=V(\theta,\phi)\omega_0J_zV^\dagger(\theta,\phi),
\end{align}
where $R(\theta,\phi)=\me^{-\mi\phi J_z}\me^{-\mi\theta J_y}$ and $V(\theta,\phi)=R(\theta,\phi)\me^{\mi\phi J_z}$.
The energy eigenstates can be constructed by the spin eigenstates $|jm\rangle$ of $J_z$:
\begin{align}\label{e20}
|\psi_m(\theta,\phi)\rangle=V(\theta,\phi)|jm\rangle=\me^{\mi m\phi}R(\theta,\phi)|jm\rangle.
\end{align}
where $m=-j,-j+1,\cdots,j-1,j$.
The details can be found in Appendix \ref{appa}. When the ensemble of spin-$j$ particles is in thermal equilibrium with temperature $T$, the canonical-ensemble density matrix is given by $\rho=\frac{1}{Z}\me^{-\beta H}$.

\section{Uhlmann process of spin-$j$ systems}\label{sec.3}


\subsection{General expressions}
Here we derive the general expression of the Uhlmann connection of generic spin-$j$ systems with the Hamiltonian given by Eq.~(\ref{sjH1}). We emphasize that the system needs to be in contact with a heat bath to stay in equilibrium due to the complications of the Uhlmann process with time evolution~\cite{ourPRB20}.
The commutator in Eq.~(\ref{GrAU}) with temperature $T$ is
\begin{align}\label{srho}
[\dif\sqrt{\rho},\sqrt{\rho}]
=\frac{\{\dif VV^{\dagger},\me^{-\beta H}\}}{Z}+\frac{2\me^{-\frac{\beta H}{2}}V\dif V^{\dagger}\me^{-\frac{\beta H}{2}}}{Z} .
\end{align}
Here $\{\cdot,\cdot\}$ denotes the anti-commutator, and the relation $\dif VV^\dagger+V\dif V^\dagger=0$ has been applied.
The details can be found in the Appendix.
Plugging $\lambda_{m}=\frac{1}{Z}\me^{-\beta m\omega_{0}}$ and Eq.~(\ref{srho}) into Eq.~(\ref{GrAU}), we obtain
\begin{align}\label{AU}
A_{U}=\sum_{mn}\chi_{mn}|\psi_{m}\rangle\langle\psi_{m}|V\dif V^{\dagger}|\psi_{n}\rangle\langle\psi_{n}|.
\end{align}
Here
\begin{align}
 \chi_{mn}=\frac{\me^{-\beta m\omega_{0}}+\me^{-\beta n\omega_{0}}-2\me^{-\frac{\beta (m+n)\omega_{0}}{2}}}{\me^{-\beta m\omega_{0}}+\me^{-\beta n\omega_{0}}}.
\end{align}
and $|\psi_m\rangle$ is the simplified notation of $|\psi_m(\theta,\phi)\rangle$.
Note that $\chi_{mn}=\chi_{nm}$ and $\chi_{nn}=0$.
It can be shown that
\begin{align}\label{UdU}
 V\dif V^{\dagger}=-\mi(J_{x}\sin\phi-J_{y}\cos\phi)\dif \theta+ \mi (J_{z}-VJ_zV^\dagger)\dif \phi.
\end{align}
The proof can be found in the Appendix.

The Uhlmann connection of the spin-$j$ system has two components: $A_U=A_U^\theta\dif\theta+A_U^\phi\dif\phi$, corresponding to the first and second terms of Eq.~(\ref{UdU}), respectively. For convenience, we derive the expressions of them separately. Since the first term on the right-hand-side of Eq.~(\ref{UdU}) commutes with $V(\theta,\phi)$, Eqs.~(\ref{e20}) and (\ref{het2a}) indicate that the corresponding matrix elements are given by
\begin{eqnarray}\label{thetame}
-\langle\psi_{m}|\mi(J_{x}\sin\phi-J_{y}\cos\phi)|\psi_{n}\rangle =\me^{\mi(n-m)\phi}\langle jm|\mi J_y|jn \rangle.\notag \\
\end{eqnarray}
Substituting this into Eq.~(\ref{AU}) and using Eq.~(\ref{e20}), the $\theta$-component of the Uhlmann connection is
\begin{align}\label{AUtheta}
A_U^\theta\dif\theta=\mi\sum_{mn}\chi_{mn}R|jm\rangle\langle jm|J_y|jn\rangle\langle jn|R^\dagger\dif\theta.
\end{align}
Next, we use the relations  $J_y=\frac{J_+-J_-}{2\mi}$ and $J_\pm|jn\rangle=\sqrt{(j\mp n)(j\pm n+1)}|jn\pm1\rangle$ to get
\begin{align}\label{AUtheta2}
A_U^\theta&=\sum_{mn}\frac{\chi_{mn}}{2}R|jm\rangle\langle jn|R^\dagger \sqrt{(j-n)(j+n+1)}\delta_{m,n+1}\notag\\
&-\sum_{mn}\frac{\chi_{mn}}{2}R|jm\rangle\langle jn|R^\dagger\sqrt{(j+n)(j-n+1)}\delta_{m,n-1}.
\end{align}
Note that only the $m=n\pm 1$ terms give nonzero contributions, and $\chi_{n+1,n}=\chi_{n-1,n}=1-\text{sech}(\frac{\beta\omega_0}{2})\equiv \chi$. This means we can pull $\chi$ out of the summation in Eq.~(\ref{AUtheta2}) with the help of the delta functions. Explicitly,
\begin{eqnarray}\label{AUtheta2b}
A_U^\theta&=&\frac{\chi}{2}\sum_{mn}R|jm\rangle\langle jn|R^\dagger\left(\sqrt{(j-n)(j+n+1)}\delta_{m,n+1}\nonumber \right. \\
&&\left. +\sqrt{(j+n)(j-n+1)}\delta_{m,n-1}\right).
\end{eqnarray}
Note the result in the bracket is just $\langle jm|J_y|jn\rangle$, so we get
\begin{align}\label{AUtheta3}
A_U^\theta&=\mi\chi R\sum_m|jm\rangle\langle jm|J_y\sum_n|jn\rangle\langle jn|R^\dagger\notag\\
&=\mi\chi RJ_yR^\dagger\notag\\
&=-\mi\chi(J_{x}\sin\phi-J_{y}\cos\phi).
\end{align}
The key idea here is that  Eq.~(\ref{AUtheta}) transforms to an expression with $\chi$ pulled out of the summation since the matrix element $\langle jm|J_y|jn\rangle$ gives nonzero contribution only when $m=n\pm 1$, where $\chi_{mn}=\chi$.

To evaluate $A^\phi_U$, we need to calculate the matrix elements of $\mi (J_{z}-VJ_zV^\dagger)$ according to Eqs.~(\ref{AU}) and (\ref{UdU}). By using $\langle\psi_m|VJ_zV^\dagger|\psi_n\rangle=m\delta_{mn}$ and $\chi_{mm}=0$, one can show that the term with $\mi VJ_z V^{\dagger}$ vanishes in $A_U^\phi$.  Therefore, only the term with $iJ_z$ contributes, giving rise to
\begin{align}\label{AUphi}
A_U^\phi&=\mi\sum_{mn}\chi_{mn}V|jm\rangle\langle jm|V^\dagger J_z V|jn\rangle\langle jn|V^\dagger\notag\\
&=-\mi\sum_{mn}\chi_{mn}R|jm\rangle\langle jm|J_x\sin\theta|jn\rangle\langle jn|R^\dagger,
\end{align}
where we have used Eq.~(\ref{het1}) in the second line and $\chi_{mn}\delta_{mn}=0$ in the last line. Since $J_x=\frac{J_++J_-}{2}$, the matrix element $\langle jm|J_x|jn\rangle$ gives nonzero contribution only when $m=n\pm 1$, where $\chi_{mn}=\chi$.  By a derivation similar to that of $A^\theta_U$, we can replace $\chi_{mn}$ by $\chi$ in the last line of Eq.~(\ref{AUphi}). Thus,
\begin{align}\label{AUphi2}
A_U^\phi\dif\phi
&=-\mi\sum_{mn}\chi R|jm\rangle\langle jm|J_x \sin\theta|jn\rangle\langle jn|R^\dagger\dif\phi\notag\\
&=-\mi \chi R J_x R^\dagger\sin\theta\dif\phi\notag\\
&=-\mi\chi\left[(J_x\cos\phi+J_y\sin\phi)\cos\theta-J_z\sin\theta\right]\sin\theta\dif\phi\notag\\
&=-\mi\frac{\chi}{\omega_0}H(\theta+\frac{\pi}{2},\phi)\sin\theta\dif\phi,
\end{align}
where Eqs.~(\ref{het2a}) and (\ref{het1a}) have been used.
In conclusion, the Uhlmann connection of the spin-$j$ system is
\begin{align}\label{AU2}
A_U&=-\mi\chi(J_{x}\sin\phi-J_{y}\cos\phi)\dif\theta\notag\\
&-\mi\frac{\chi}{\omega_0}H(\theta+\frac{\pi}{2},\phi)\sin\theta\dif\phi.
\end{align}
The Uhlmann curvature is given by
\begin{equation}\label{FU}
F_U=\dif A_U+A_U\wedge A_U.
\end{equation}
The Chern number associated with the Uhlmann connection is
$\text{Ch}_U=\frac{\mi}{2\pi}\mathlarger{\int}\text{Tr}F_U$.
As summarized in the Appendix, the Chern number associated with the Uhlmann connection vanishes.
This is consistent with the fact that the Uhlmann bundle is a trivial one~\cite{DiehlPRB15}.

We remark that the previous discussions assumes $j>0$ since $J_x$, $J_y$, and $J_z$ only have nontrivial representations if $j>0$. However, the $j=0$ system has only a single state. Thus, the $j=0$ system is in a pure state. As shown in the Appendix, both
$A_U=0$ and $\theta_U=0$ for this particular case.

\subsection{Evaluation of Loschmidt Amplitude}\label{sec.4}
When evaluating the Loschmidt amplitude according to Eq.~(\ref{GrU0}), we need to evaluate the path-ordered integral over a loop parameterized by $t$ in the parameter space. If $A_U$ is a diagonal matrix or a constant matrix, the path-ordering operation is automatically satisfied since all the $A_U(t)$ at different $t$ commute with each other. Thus, an explicit expression of $\mathcal{G}^U_\rho$ can be obtained for some special cases undergoing Uhlmann processes. Refs.~\cite{ViyuelaPRL14,ourPRB20} show the Uhlmann phase of a two-level system traversing a great circle in the parameter space. For the spin-$j$ system, this can be achieved by choosing a suitable loop on the parameter space $S^2$. In the following, we show the results following a circle of longitude and the circle of latitude at $\theta=\frac{\pi}{2}$, i.e. the equator.

\subsubsection{Circle of longitude}
We first consider the system traversing a great circle of fixed longitude $\phi$ in the parameter space. Hence, $\dif \phi=0$, and Eq.~(\ref{AU2}) indicates
\begin{align}\label{AU3}
\oint A_U&=-\mi\chi(J_{x}\sin\phi-J_{y}\cos\phi)\oint \dif\theta\notag\\
&=2\pi\mi\chi\mathrm{e}^{-\mathrm{i} \phi J z} J_{y} \mathrm{e}^{\mathrm{i} \phi J_{z}}\Omega,
\end{align}
where $\Omega\equiv \frac{1}{2\pi}\mathlarger{\oint} \dif\theta$ denotes the winding number along the circle of longitude during the Uhlmann process.
We further assume the Uhlmann process starts from the north pole with $\theta(0)=0$. Thus, the initial Hamiltonian is $H(0)=\omega_0J_z$, and the corresponding density matrix is
\begin{align}\label{r0}
\rho(0)=\frac{1}{Z(0)}\me^{-\beta\omega_0J_z}.
\end{align}
Substitute Eqs. (\ref{AU3}) and (\ref{r0}) into Eq.~(\ref{GrU0}), we have
\begin{align}\label{G1}
\mathcal{G}^U_{\theta}(T)&=\sum_{m=-j}^{j}\frac{ \mathrm{e}^{-\beta \omega_{0} m}}{Z(0)}\langle j m|\mathrm{e}^{-2 \pi \Omega \chi \mathrm{i} \mathrm{e}^{-\mathrm{i} \phi J z} J_{y} \mathrm{e}^{\mathrm{i} \phi J_{z}}}| j m\rangle \notag\\
&=\sum_{m=-j}^{j}\frac{ \mathrm{e}^{-\beta \omega_{0} m}}{Z(0)}\langle j m|\mathrm{e}^{-\mathrm{i} \phi J_{z}} \mathrm{e}^{-2 \pi \Omega \chi \mathrm{i} J_{y}} \mathrm{e}^{\mathrm{i} \phi J_{z}}| j m\rangle\notag\\
&=\sum_{m=-j}^{j}\frac{ \mathrm{e}^{-\beta \omega_{0} m}}{Z(0)}d^j_{mm}(2\pi\Omega\chi).
\end{align}
Here $d^j_{mm'}(\Theta)=\langle jm|\me^{-\mi\Theta J_y}|jm'\rangle$ is the Wigner $d$-function. Interestingly, the result is independent of the longitude, which will be discussed later.

\subsubsection{Equator}
In this situation, the system traverses the equator in the parameter space with the latitude fixed at $\theta=\frac{\pi}{2}$. Thus, $\dif\theta=0$. According to Eqs.~(\ref{AUphi2}) and (\ref{AU2}), $A_U$ becomes
\begin{align}\label{AU4}
A_U=\mi\chi J_z\dif\phi,
\end{align}
which further implies $\mathlarger{\oint}A_U=2\pi\mi\chi J_z\Omega$. Here $\Omega\equiv \frac{1}{2\pi}\mathlarger{\oint} \dif\phi$ is the winding number along the equator during the Uhlmann process. We assume the Uhlmann process starts from the point $(\theta=\frac{\pi}{2},\phi=0)$. Using Eq.~(\ref{sjH1}), the density matrix is given by
\begin{align}\label{r1}
\rho(0)&=\frac{1}{Z(0)}\me^{-\beta\omega_0\me^{-\mi\frac{\pi}{2}J_y}J_z\me^{\mi\frac{\pi}{2}J_y}}\notag\\
&=\frac{1}{Z(0)}\me^{-\mi\frac{\pi}{2}J_y}\me^{-\beta\omega_0J_z}\me^{\mi\frac{\pi}{2}J_y}.
\end{align}
Substitute this into Eq.~(\ref{GrU0}), the Loschmidt amplitude is
\begin{align}\label{G2}
&\mathcal{G}^U_\phi(T)=\frac{1}{Z(0)}\text{Tr}\left(\me^{-\mi\frac{\pi}{2}J_y}\me^{-\beta\omega_0J_z}\me^{\mi\frac{\pi}{2}J_y}\me^{-2\pi\Omega\chi\mi J_z}\right)\notag\\
&=\frac{1}{Z(0)}\text{Tr}\left[\me^{-\beta\omega_0J_z}R^\dagger\left(\frac{\pi}{2},-\frac{\pi}{2}\right)\me^{-2\pi\Omega\chi\mi J_z}R\left(\frac{\pi}{2},-\frac{\pi}{2}\right)\right]\notag\\
&=\sum_{m=-j}^{j}\frac{ \mathrm{e}^{-\beta \omega_{0} m}}{Z(0)}d^j_{mm}(2\pi\Omega\chi),
\end{align}
where $R$ is the unitary transformation shown in Eq.~(\ref{sjH1}).
Therefore, $\mathcal{G}^U_{\theta}(T)=\mathcal{G}^U_\phi(T)$. This is not surprising since the circle of latitude and the equator are both great circles on $S^2$. One may conjecture that the Loschmidt amplitude along any great circle on the parameter space $S^2$ has the same expression, of which a mathematical proof is not the focus of this paper. Based on the discussion, we will use $\mathcal{G}^U$ to denote the Loschmidt amplitude hereafter since it represents the result from a great circle.

\section{Examples}\label{sec.5}
\subsection{$j=\frac{1}{2}$}

\begin{figure}[th]
\centering
\includegraphics[width=3.3in]{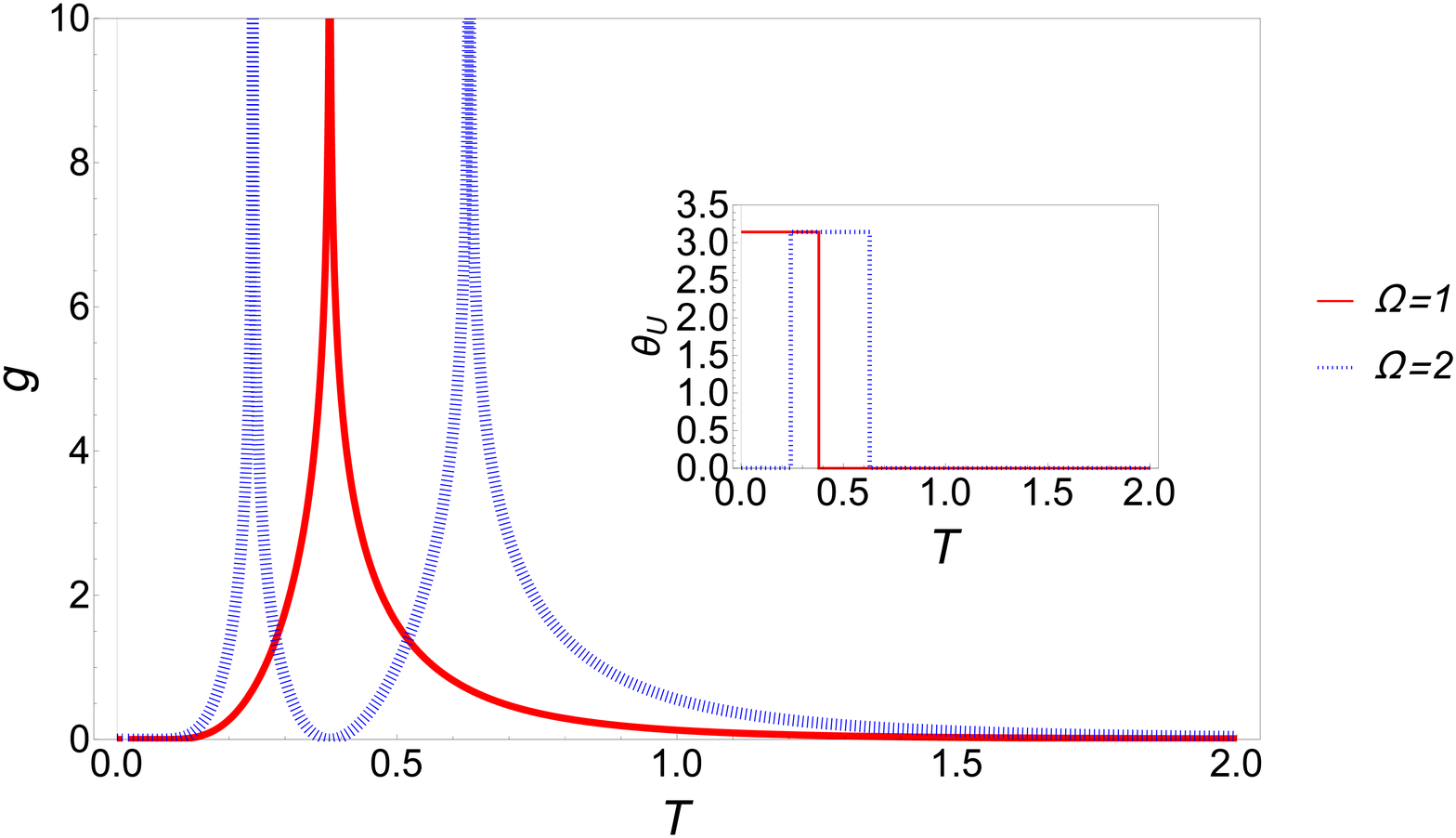}
\includegraphics[width=3.1in]{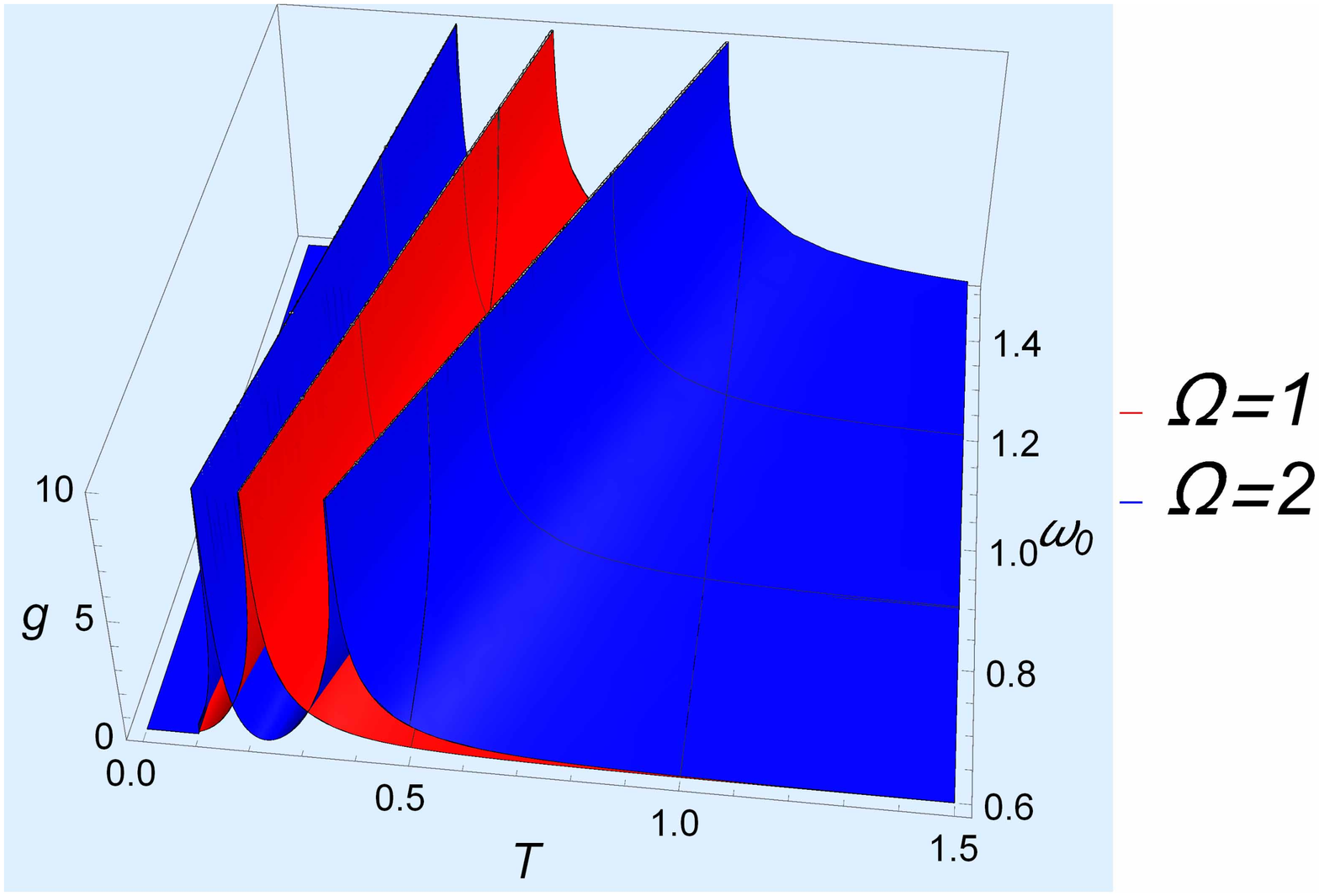}
\caption{(Top panel) The geometrical generating function $g$ vs. $T$ for the spin-$\frac{1}{2}$ system with $\omega_0=1.0$. The red solid and blue dotted lines correspond to $\Omega=1,2$, respectively. The TQPTs occur at the diverging peaks, corresponding to $T^*$s. The inset plots the Uhlmann phase vs. $T$, showing a jump at each $T^*$. (Bottom panel) The geometrical generating function $g$ vs. $T$ and $\omega_0$ for $\Omega=1, 2$, respectively.}
\label{Fig1}
\end{figure}

After deriving the generic formula of a spin-$j$ system in an Uhlmann process, we will use explicit examples to analyze the topological properties of an Uhlmann process.
We first consider the $j=\frac{1}{2}$ system, where the angular momentum is $J_i=\frac{1}{2}\sigma_i$ with $i=x,y,z$ and $\sigma_i$ being the Pauli matrices. The two energy levels are $|\frac{1}{2},\pm\frac{1}{2}\rangle=\begin{pmatrix}1 \\ 0\end{pmatrix},\begin{pmatrix}0 \\ 1\end{pmatrix}$. From Eq.~(\ref{AU2}), the Uhlmann connection is
\begin{align}\label{A0.5}
A_U=\frac{\chi}{2}\begin{pmatrix}0 & \me^{-\mi\phi}\\-\me^{\mi\phi} & 0\end{pmatrix}\dif\theta-
\frac{\mi\chi}{2}\begin{pmatrix}-\sin\theta & \cos\theta\me^{-\mi\phi}\\\cos\theta \me^{\mi\phi} & \sin\theta\end{pmatrix}\dif\phi.
\end{align}
If the Uhlmann process corresponds to a circle of longitude or the equator in the parameter space, $A_U$ is proportional to a constant matrix. Using $Z(0)=\me^{-\frac{1}{2}\beta\omega_0}+\me^{\frac{1}{2}\beta\omega_0}$ and $d^{\frac{1}{2}}_{-\frac{1}{2}-\frac{1}{2}}(2\pi\Omega\chi)=d^{\frac{1}{2}}_{\frac{1}{2}\frac{1}{2}}(2\pi\Omega\chi)=\cos(\pi\Omega\chi)$, either Eq.~(\ref{G1}) or (\ref{G2}) leads to
\begin{align}\label{G0.5}
\mathcal{G}^U(T)=\cos(\pi\Omega)\cos\left(\pi\Omega\text{sech}\frac{\beta\omega_0}{2}\right).
\end{align}
The expression is quite similar to that of the two-band models~\cite{ViyuelaPRL14,ourPRB20} since they are both two-level systems. However, the energy spectrum of the $j=1/2$ paramagnet is independent of the parameters spanning the parameter space while the band structure of a two-band model explicitly depends on the crystal momentum. Moreover, the spin $1/2$ model allows the winding number in the parameter space to be any integer. In contrast, Ref.~\cite{ViyuelaPRL14} considered periodic two-band systems with the crystal momentum in the Brillouin zone, and the winding number is at most $1$.
Similarly, Ref.~\cite{Galindo21} considers only the case with $\Omega=1$. As we will show shortly, higher winding numbers will introduce interesting physics at finite temperatures.

The low- and high- temperature limits can be analyzed by Eq.~(\ref{G0.5}). In the zero-temperature limit, $T\rightarrow 0$, $\text{sech}\left(\frac{\beta\omega_0}{2}\right)\rightarrow 0$ since $\beta\rightarrow\infty$. Thus, $\mathcal{G}^U(T\rightarrow 0)=\cos(\pi\Omega)=(-1)^\Omega$, which implies $\theta_U=0$ (topologically trivial) if $\Omega$ is even and $\theta_U=\pi$ (topologically nontrivial) if $\Omega$ is odd. Here the topology refers to that of the horizontal lift, whose pictorial description will be given in Sec.~\ref{app:fb}. In the infinite-temperature limit, $T\rightarrow\infty$, $\text{sech}\left(\frac{\beta\omega_0}{2}\right)=1$ since $\beta\rightarrow0$, which implies $\mathcal{G}^U(T\rightarrow \infty)=\cos^2(\pi\Omega)=1$. Thus, $\theta_U=0$ and the system is always topologically trivial at infinitely high temperature.

A TQPT occurs at temperature $T^*$ when the Loschmidt amplitude vanishes, $\mathcal{G}^U(T^*)=0$, accompanied by a jump of the Uhlmann phase~\cite{ourPRB20b}. For the $j=1/2$ system, Eq.~(\ref{G0.5}) implies
\begin{align}\label{T*0.5a}
\pi\Omega\text{sech}\left(\frac{\omega_0}{2T^*}\right)=\left(n+\frac{1}{2}\right)\pi
\end{align}
or equivalently
\begin{align}\label{T*0.5b}
\cosh\left(\frac{\omega_0}{2T^*}\right)=\frac{\Omega}{n+\frac{1}{2}}.
\end{align}
Here $n$ is an integer. From the above expression, we get
\begin{align}\label{T*0.5}
T^*=\frac{\omega_0}{2\ln\left(\frac{\Omega}{n+\frac{1}{2}}+\sqrt{\left(\frac{\Omega}{n+\frac{1}{2}}\right)^2-1}\right)}.
\end{align}
We only consider the situation with $\Omega>0$, which requires $n>0$ to ensure $T^*>0$. Another
premise of Eq.~(\ref{T*0.5}) is $\frac{\Omega}{n+\frac{1}{2}}\ge1$, i.e., $n=0,1,\cdots \Omega-1$. This indicates that the number of positive solutions of $T^*$ is equivalent to $\Omega$, which is not addressed in Ref.~\cite{Galindo21} due to the limitation of $\Omega=1$ there. Therefore, the winding number in the parameter space decides how many TQPTs the $j=1/2$ systems will go through as $T$ increases.

To clearly see the topological features of the system, we show the numerical results in Figure~\ref{Fig1}. If $\Omega=0$, $\mathcal{G}^U(T)=1$ at any temperature, showing only a topologically trivial phase. In contrast, we plot the geometrical generating function $g$ as a function of both $T$ and $\omega_0$ with $\Omega=1$ and $2$ in the bottom panel. The diverging peaks indeed appear at the phase transition temperatures $T^*$, indicating the locations of the TQPTs.
In the top panel, we show the cross-section of the bottom panel at $\omega_0=1.0$. The inset shows the Uhlmann phase $\theta_U$ as a function of $T$. If $\Omega=1$ (denoted by the red line), the system is in a topological phase ($\theta_U=\pi$) at low temperatures. As the temperature increases, the system transits to the topologically trivial phase at $T^*$ and $\theta_U$ jumps to zero.

If $\Omega=2$ (denoted by the blue dotted line), the system is in a topologically trivial phase ($\theta_U=0$) at low temperatures. As the temperature increases, the system undergoes two phase transitions at different values of $T^*$: It first transits to a topologically nontrivial phase at the first $T^*$ and then becomes topologically trivial when crossing the second $T^*$. $\theta_U$ jumps at these $T^*$'s. Importantly, the topologically nontrivial phase is sandwiched between two topologically trivial phases and only survive at finite temperatures. Although the $\Omega=1$ case gives the impression that temperature destroys topological properties, we see the $\Omega=2$ case gives an example that a topological regime is only possible at finite temperatures. For the spin-$1/2$ system, the Uhlmann holonomy group formed by $\mathcal{P}\me^{-\oint A_U}$ is the Z$_2$ group because the phase is defined modulo $2\pi$ and the Loschmidt amplitude given by Eq.~\eqref{G0.5} is real. In fact, the Uhlmann holonomy
group for the spin-$j$ system with arbitrary $j$ is always the Z$_2$ group since Eq.~(\ref{G1}) implies the Loschmidt amplitude is real because the partition function $Z$ and the Wigner $d$-function are both real-valued~\cite{LamQM}.

\subsection{$j=1$}
\begin{figure}[th]
\centering
\includegraphics[width=3.3in]{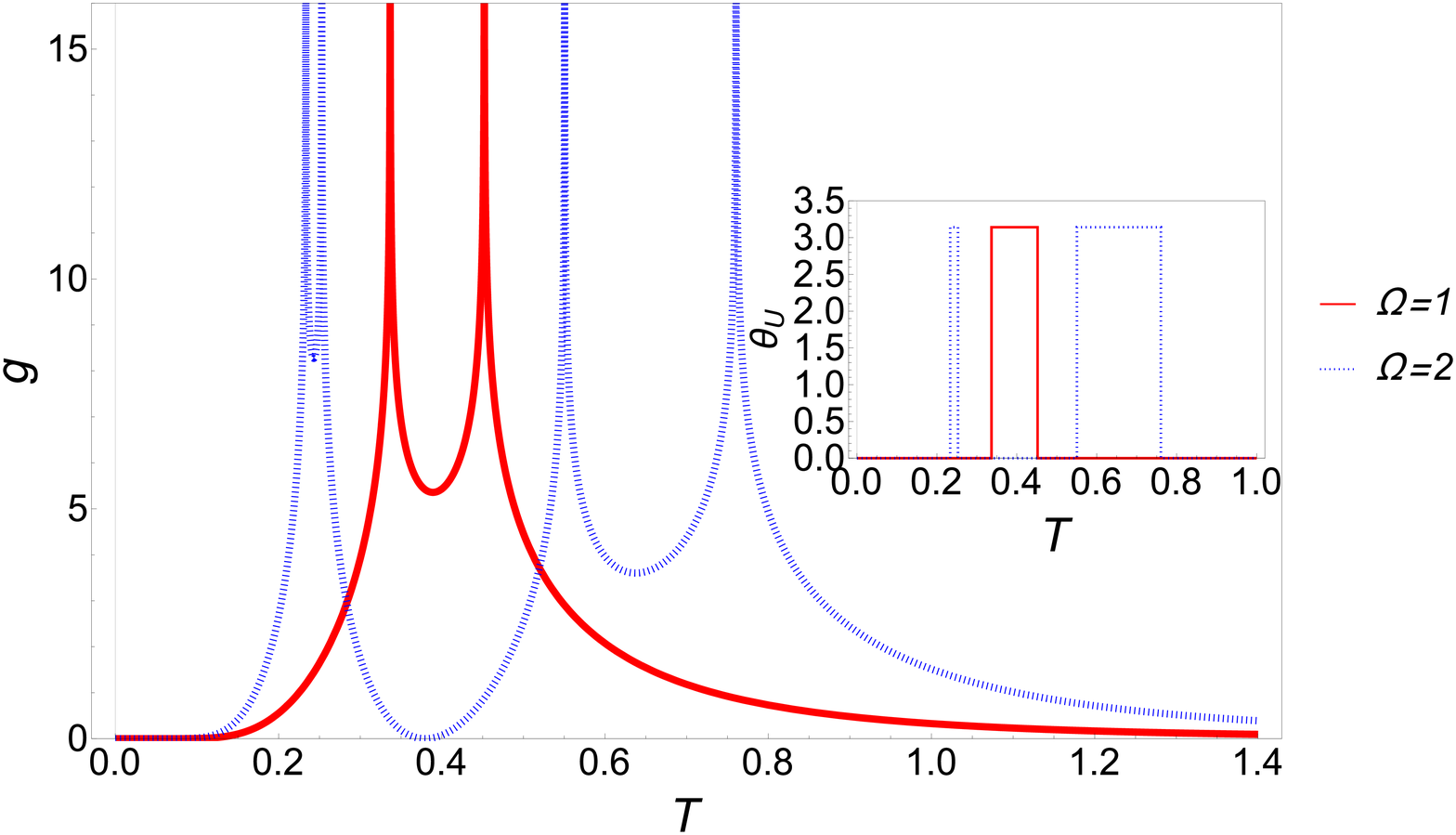}
\includegraphics[width=3.1in]{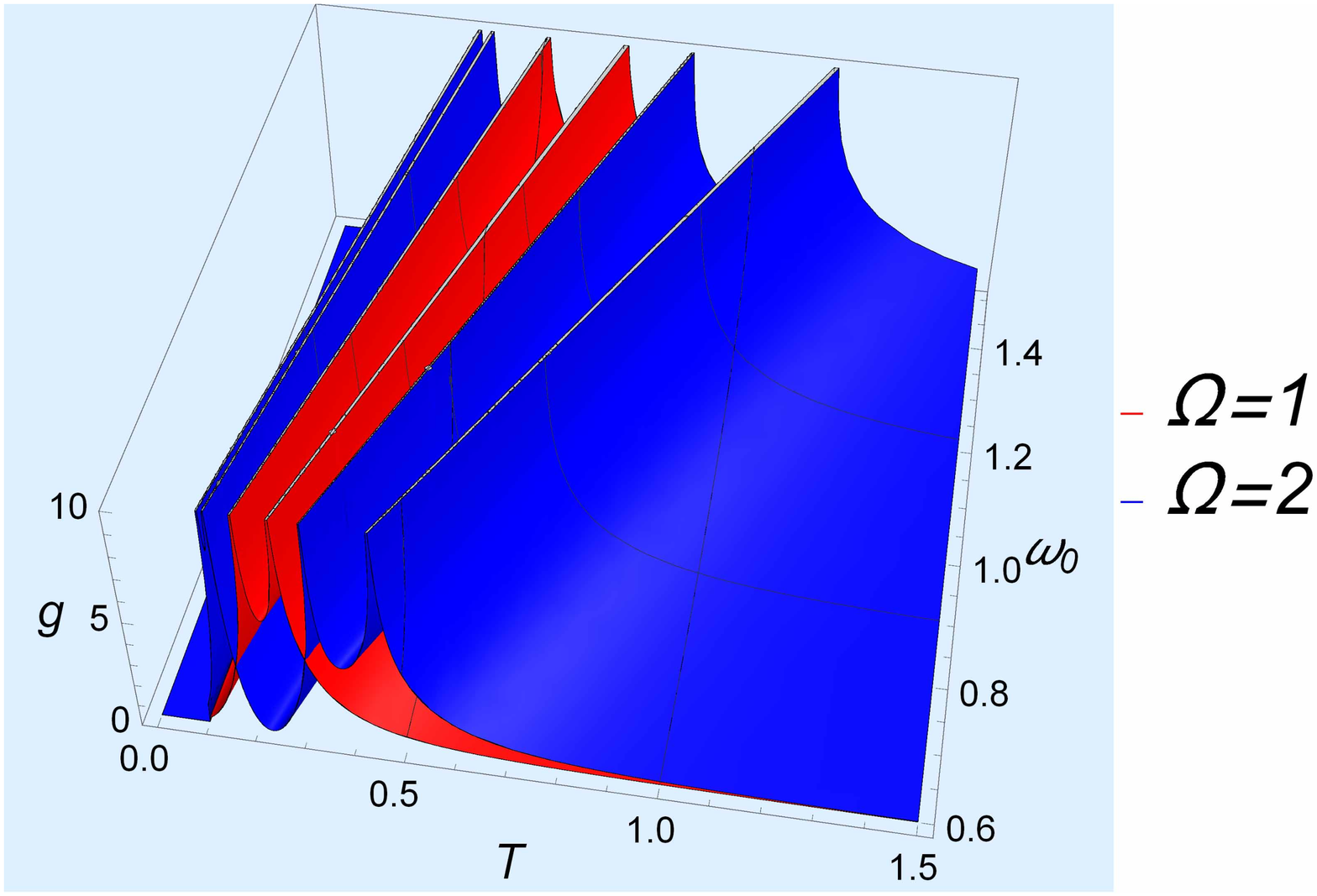}
\caption{The geometrical generating function $g$ for the spin-$1$ system as a function of $T$ with $\omega_0=1.0$ (top panel) and as a function of both $T$ and $\omega_0$ (bottom panel). The Uhlmann phase $\theta_U$ as a function of $T$ is shown in the inset of the top panel. The red and blue curves/surfaces respectively correspond to $\Omega=1,2$.  }
\label{Fig2}
\end{figure}

We now turn to the more complicated spin-1 system. The matrix representations of the three components of the angular momentum in units of $\hbar$ are $J_x=\frac{1}{\sqrt{2}}\begin{pmatrix} 0 & 1 & 0\\ 1 & 0 & 1\\ 0 & 1 & 0\end{pmatrix}$, $J_y=\frac{1}{\sqrt{2}\mi}\begin{pmatrix} 0 & 1 & 0\\ -1 & 0 & 1\\ 0 & -1 & 0\end{pmatrix}$ and $J_z=\begin{pmatrix} 1 & 0 & 0\\ 0 & 0 & 0\\ 0 & 0 & -1\end{pmatrix}$. Following Eq.~(\ref{AU2}), we obtain
\begin{align}\label{AUj1}
A_U&=\frac{\chi}{\sqrt{2}}\begin{pmatrix} 0 & \me^{-\mi\phi} & 0\\ -\me^{\mi\phi} & 0 & \me^{-\mi\phi}\\ 0 & -\me^{\mi\phi} & 0\end{pmatrix}\dif\theta\notag\\&-
\frac{\mi\chi}{\sqrt{2}}\begin{pmatrix} -\sqrt{2}\sin\theta & \cos\theta\me^{-\mi\phi} & 0\\ \cos\theta\me^{\mi\phi} & 0 & \cos\theta\me^{-\mi\phi}\\ 0 & \cos\theta\me^{\mi\phi} & \sqrt{2}\sin\theta\end{pmatrix}\dif\phi.
\end{align}
Similarly, the analytical result of the Loschmidt amplitude can be obtained by using the expressions of Wigner's $d$-functions if the Uhlmann process corresponds to a circle of longitude or the equator. Explicitly,
\begin{align}\label{Gj1}
\mathcal{G}^U(T)=\frac{1}{Z(0)}&\bigg\{\cosh(\beta\omega_0)\left[1+\cos\left(2\pi\Omega\text{sech}\frac{\beta\omega_0}{2}\right)\right]\notag\\&+\cos\left(2\pi\Omega\text{sech}\frac{\beta\omega_0}{2}\right)\bigg\},
\end{align}
where $Z(0)=1+2\cosh(\beta\omega_0)$.

Again, $\Omega=0$ leads to $\mathcal{G}^U(T)=1$, so the system is topologically trivial at any temperature. Hence, we only consider the situations with $\Omega\ge 1$. Moreover, Eq.~(\ref{Gj1}) has the following limits: $\mathcal{G}^U(T\rightarrow0)=1=\mathcal{G}^U(T\rightarrow\infty)$, indicating the system is topologically trivial at low and high temperatures, which is different from the $j=1/2$ system allowing topologically non-trivial phase at low temperatures. To check if the $j=1$ system has finite-temperature topological regimes, we numerically analyze $\mathcal{G}^U(T)$ and locate its zeros. We visualize our numerical results in Figure.~\ref{Fig2}, where the geometrical generating function $g$ is plotted as a function of $T$ and $\omega_0$ for $\Omega=1$ and $2$ in the bottom panel. We follow the same convention as Fig.~\ref{Fig1}. One clearly sees that $g$ diverges at several values of $T^*$. The divergence corresponds to a zero of $\mathcal{G}^U$, indicating the occurrence of a finite-temperature TQPT. At each $T^*$, the Uhlmann phase jumps, as shown in the inset of Fig.~\ref{Fig2}

Whether a phase is topological or trivial can be inferred from the quantized Uhlmann phase $\theta_U$. The topologically trivial regimes with vanishing Uhlmann phase at low and high temperatures confirm our previous analysis. Different from the $j=\frac{1}{2}$ case, all the topological regimes of the $j=1$ system are at finite temperatures. There are $2\Omega$ TQPT points in the $j=1$ system when $T$ increase with a fixed value of $\omega_0$, as one can see on Fig.~\ref{Fig2}. Those TQPTs indicate there are $\Omega$-numbered finite-temperature topological regimes. If $\Omega=2$, the system goes through a sequence of trivial, nontrivial, trivial, nontrivial and trivial phases, showing two topological regimes as temperature increases from $0$ to $\infty$ in an Uhlmann process. The Uhlmann holonomy group is the Z$_2$ group, as one can check that the Loschmidt amplitude given by Eq.~\eqref{Gj1} is real.

\subsection{Absence of TQPT in Uhlmann process at infinite temperature}
Starting with an initial state at infinite temperature, a quench process may exhibit DQPTs as the system evolves out of equilibrium~\cite{DTQPT18,ourPRB20b}. In contrast, here we show that there is no infinite-temperature TQPT from an Uhlmann process because the density matrix needs to be in equilibrium at infinite temperature in the cyclic process. The following proof applies to any Uhlmann process, not just those of the spin-$j$ systems. Since the proof is more straightforward if the fiber-bundle language is used, we will give a brief overview of the description first.

\subsubsection{Fiber-bundle description}\label{app:fb}
Following Refs.~\cite{Uhlmann89,ourPRB20}, we consider a $n$-level system and introduce an associated fiber bundle $(E,\pi,Q,F,\text{U}(n))$. Here $E$ is the total space and $Q$ is the base space spanned by the full-rank density matrix $\rho$. $\pi$ is the projection that acts as $\pi:E\rightarrow Q$. Explicitly,
 \begin{align}
\pi(W)=WW^\dagger=\rho.
\end{align}
A smooth map $\sigma$: $Q\rightarrow E$ that satisfies $\pi\circ\sigma=1_Q$ is called a section, where $1_Q$ is the identity map on $Q$.
$F$ is the fiber spanned by the amplitudes, i.e., it is isomorphic to $H_W$ described below Eq.~\eqref{w2}. U$(n)$ is the structure group formed by the elements that act on the fiber.
To understand the local structure of the fiber bundle, we consider a set of open coverings $\{Q_i\}$ of the base space $Q$. For an arbitrary density matrix $\rho\in Q_i$, a local trivialization $\phi_i$: $Q_i\times F\rightarrow \pi^{-1}(Q_i)$ satisfies $\phi_i^{-1}(\pi^{-1}(\rho))=(\rho, W)$, where $\rho=WW^\dagger$. Hence, the fiber $F_\rho$ above the base point $\rho$ is spanned by all amplitudes satisfying $W=\sqrt{\rho}U$.
It can be shown that the fiber bundle is in fact a principle bundle since $F$ is diffeomorphic to U$(n)$~\cite{Asorey19}. Moreover, it is also a trivial bundle since it admits a global section~\cite{DiehlPRB15} $\sigma(\rho)=\sqrt{\rho}$.

The parallel transport of the amplitude has been discussed previously, but it can also be described in the fiber-bundle language.
When a physical system varies along a curve in the parameter space, $\gamma(t)$: $[0,\tau]\rightarrow M$, the density matrix $\rho(t)\equiv \rho(\gamma(t))$ varies along a corresponding curve $\mathcal{C}$ in $Q$. By using the local trivialization $\phi_i^{-1}(\pi^{-1}(\rho(t)))=(\rho(t), W(t))$, it can be shown that the associated amplitudes must also change continuously along a certain curve $\tilde{\mathcal{C}}$ in $E$. If $W(t)$ is parallel transported along $\tilde{\mathcal{C}}$ satisfying Eq.~(\ref{Wpt}), it is equivalent to say that $\tilde{\mathcal{C}}$ is a horizontal lift of $\mathcal{C}$. This further requires the tangent vector $\tilde{X}$ of the curve $\tilde{\gamma}$ to be a horizontal vector, which belongs to the horizontal subspace of the tangent bundle $TE$. Thus, an Ehresmann connection $\omega$ on $E$ is needed to separate $TE$ into the horizontal and vertical spaces as $TE=HE\oplus VE$, and the horizontality condition is
 \begin{align}\label{X}
\omega(\tilde{X})=0.
\end{align}
Moreover, the section $\sigma: Q\rightarrow E$ induces a pullback of $\omega$ as $A_U=\sigma^*\omega$, which is a connection on the base space $Q$. If the horizontality condition is satisfied, $A_U$ is the Uhlmann connection. If the Uhlmann connection is defined, the fiber bundle is referred to as the Uhlmann bundle.  The horizontality condition (\ref{X}) can be cast into the form
\begin{align}\label{AUEn}
A_U(X)=-\dif U(\tilde{X})U^\dagger.
\end{align}
Here $U$ is the phase factor in Eq.~(\ref{w2}). $X=\pi_*(\tilde{X})$ is the tangent vector of the curve $\mathcal{C}$ in $Q$, where $\pi_*$ is the push-forward induced by the projection $\pi$.
If the curve $\mathcal{C}$ is closed, i.e. the related process is cyclic, then it corresponds to the Uhlmann process.
More details can be found in Ref.~\cite{ourPRB20}.

The Uhlmann phase reveals the Uhlmann holonomy given by $W(0)$ and $W(1)$, as illustrated in Fig.~\ref{Fig3}. Since the phase is defined modulo $2\pi$, one can see that a $\pi$ jump of the Uhlmann phase signifies a change of the topology of the horizontal-lift curve. 

\subsubsection{Proof of no TQPT in Uhlmann process at infinite temperature}
\begin{figure}[th]
\centering
\includegraphics[width=3.3in]{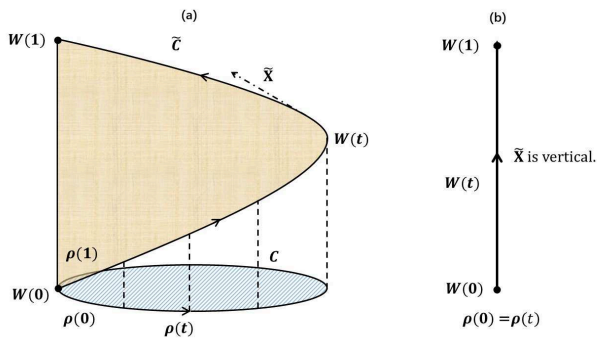}
\caption{(a) A pictorial illustration of parallel-transport of an amplitude. The Uhlmann holonomy reflects the change along the curve $\tilde{\mathcal{C}}$, which is the horizontal lift of the closed curve $\mathcal{C}$ along which the density matrix evolves. The tangent vector $\tilde{X}$ must be horizontal. Here $W(0)\neq W(1)$ are both amplitudes of the same density matrix $\rho(0)=\rho(1)$, meaning they belong to the same fiber at $\rho(0)$. The dashed lines denote the projection. (b) Illustration of the infinite-temperature limit. The curve $\mathcal{C}$ shrinks to a single point, so its horizontal lift $\tilde{\mathcal{C}}$ must also shrink to a single point. Otherwise, $W(0)\neq W(1)$, and the tangent vector $\tilde{X}$ of $\tilde{\mathcal{C}}$ is vertical since $\tilde{\mathcal{C}}$ lies solely in a single fiber. This contradicts the horizontal-lift condition of $\tilde{\mathcal{C}}$.}
\label{Fig3}
\end{figure}

At infinitely high temperature, the density matrix is proportional to the identity operator, i.e., $\rho=\frac{1}{\dim\rho}\hat{1}$, no matter how the system changes along any curve in the parameter space.
Therefore, the curve $\mathcal{C}$ in $Q$ along which $\rho$ varies becomes trivial, equivalent to a single point. Thus, the horizontal lift $\tilde{\mathcal{C}}$ also becomes trivial. This can be proven by reductio ad absurdum. If $\tilde{\mathcal{C}}$ is nontrivial, it must be a curve solely belonging to a single fiber at the point $\rho=\frac{1}{\dim\rho}\hat{1}$. Thus, its tangent vector $\tilde{X}$ must be vertical, which contradicts the horizontal condition for parallel transport. Explicitly, the amplitude of $\rho=\frac{1}{\dim\rho}\hat{1}$ can be expressed as  $W(t)=\frac{1}{\sqrt{\dim\rho}}\me^{t u}$, where $u$ is an anti-Hermitian matrix: $u=-u^\dagger$. Thus, $\dot{W}=Wu$. Since $W$ is parallel-transported along $\tilde{\mathcal{C}}$, it satisfies Eq.~(\ref{Wpt}), which then leads to
\begin{align}\label{u}
-uW^\dagger W=W^\dagger Wu.
\end{align}
Since $W^\dagger W=WW^\dagger=\frac{1}{\dim\rho}\hat{1}$, Eq.~(\ref{u}) implies that $u=0$. Thus, $W(t)=\frac{1}{\sqrt{\dim\rho}}\hat{1}$, which means $\tilde{\mathcal{C}}$ is also trivial. This further leads to the Uhlmann fidelity $\mathcal{G}^U(T\rightarrow \infty)=\text{Tr}(W^\dagger(0)W(1))=1$. As a consequence, no TQPT from an Uhlmann process can occur at infinite temperature.
A pictorial description of the absence of any TQPT from a Uhlmann process at infinite temperature is given in Figure~\ref{Fig3} (b).

Before discussing experimental implications, we explain the topological regimes at finite temperatures in Figs.~\ref{Fig1} and \ref{Fig2}. Since the Loschmidt amplitude for the spin $j$ systems is real, the Uhlmann phase can only be $0$ (trivial) or $\pi$ (non-trivial) modulo $2\pi$. Increasing the temperature can lead to more twists of the horizontal lift shown in Fig.~\ref{Fig3}, causing the Uhlmann phase to jump. However, the Uhlmann phase is always trivial at infinite temperature, as proven above. Therefore, if the low-temperature regime is trivial with $\theta_U=0$ due to higher winding number or spin, as shown in the spin $1/2$ case with $\Omega=2$ or the spin $1$ case with $\Omega=1, 2$, there must be an even number of jumps of $\theta_U$ between $T=0$ and $T\rightarrow \infty$. Those jumps signify the TQPTs and enclose topologically nontrivial regimes at finite temperatures. The finite-temperature topological regimes of the spin-$j$ systems studied here thus represent consecutive jumps of the Uhlmann holonomy shown in Fig.~\ref{Fig3} above a trivial ground state as temperature increases.

\section{Implication for experiment}\label{sec.6}
\subsection{Parallel transport of purified states}
We have analyzed the topological property of a spin-$j$ system in an Uhlmann process. It is important to realize the Uhlmann process and measure the Uhlmann phase using natural or engineered systems. Here we discuss how the predictions may be verified in future experiments. Before explicit experimental protocols are discussed, however, some fundamental problems need to be addressed.
The first problem is to suitably characterize the amplitude of a density matrix because there is no one-to-one correspondence between a mixed quantum state and an amplitude, given $W=\sqrt{\rho}U$ and the arbitrariness of $U$. Moreover, a density matrix may not uniquely corresponds to a mixed quantum state, either. However, Eq.~(\ref{w2}) indicates that a given amplitude can be represented by a purified state, which is formally a pure quantum state. Thus, one can employ an ancilla state entangled with the system state to form and manipulate a purified state. Recently, Ref.~\cite{npj18} has shed light on this issue by using quantum-computer simulations to analyze two-band systems.

The second problem is how to physically perform the parallel transport of an amplitude following Eq.~(\ref{Wpt}). Integrating both sides of Eq.~(\ref{Wpt}) leads to the parallel condition (\ref{Wpt1}), which involves a matrix product of two amplitudes in $H_W$.
Unfortunately, there is no operation between two purified states in $\mathcal{H}\otimes \mathcal{H}$ corresponding to such a matrix product in $H_W$, although the inner product in the former is isomorphic to the Hilbert-Schmidt product in the latter. Thus, by using the purified-state representation of the amplitude, we have to employ a weaker condition for parallel transport. By taking the trace of both sides of Eq.~(\ref{Wpt1}), we get
\begin{align}\label{dm0}
\langle W_1|W_2\rangle=\langle W_2|W_1\rangle, \implies \text{Im}\langle W_1|W_2\rangle=0.
\end{align}
This condition minimizes the Fubini-Study distance between $|W_1\rangle$ and $|W_2\rangle$ just as the condition $(W_1,W_2)=(W_2,W_1)$ minimizes the Hilbert-Schmidt
distance between $W_1$ and $W_2$~\cite{Uhlmann86}. If the purified state is transported along a curve parametrized by $t$ with $|W_1\rangle=|W(t)\rangle$ and $|W_2\rangle=|W(t+\dif t)\rangle$, the differential form of Eq.~(\ref{dm0}) is given by
\begin{align}\label{dm0a}
\text{Im}\langle W(t)|\frac{\dif}{\dif t}|W(t)\rangle=0.
\end{align}
This is the same condition as that for an adiabatic process during which a pure quantum state obtains a Berry phase~\cite{Nakahara,DiehlPRB15,ourPRB20}. This may not be surprising since the Uhlmann phase is a generalization of the Berry phase. Though the new condition (\ref{dm0a}) is weaker than the original parallel-transport condition (\ref{Wpt1}), it is experimentally realizable because constructing a purified state may sometimes be more practicable in experiments. Since Eq.~(\ref{dm0a}) only involves the purified states, the incompatibility between a dynamical process and an Uhlmann process~\cite{ourPRB20} does not apply here. Thus, $t$ can be chosen as the time, i.e., the Uhlmann process can be simulated by a suitably chosen evolution of the purified state as long as the condition (\ref{dm0a}) is respected.

\subsection{Experimental procedures}
We now turn to experimental setups and protocols for spin-$j$ systems.
Experimentally, the spin-$j$ state $|jm\rangle$ may be realized by the hyperfine states of atoms~\cite{FootBook,PethickBook}, which allows well-controlled preparation, manipulation, and measurement.
A two-level system is often referred to as a qubit. Accordingly, the spin-$j$ system might be named a qu$j$it, which has $2j+1$ components.
The Hamiltonian (\ref{sjH}) may be realized by coupling the hyperfine states with external magnetic fields. By changing the direction of the magnetic field continuously, a closed Uhlmann trajectory in the parameter space can be realized. We consider a loop of longitude with $\theta(t)$, $0\le t\le 1$.  $\theta(0)=0$ and $\theta(1)=2\pi\Omega$ at the latitude $\phi=0$, i.e. the meridian. According to Eq.~(\ref{AU2}), $A_U=-\mi\chi J_y\dif\theta$ with $\chi=1-\text{sech}(\beta\omega_0/2)$. Hence, the final phase factor is
\begin{align}\label{fpf}
U(1)=\me^{-\mathlarger{\oint} A_U}U(0)=\me^{-\mi\chi\mathlarger{\oint}\dot{\theta }\dif t J_y}U(0),
\end{align}
where $\dot{\theta}=\frac{\dif \theta(t)}{\dif t}$ and the path ordering has been fulfilled since $A_U$ at different $t$ commutes with each other.

At temperature $T$, the family of density matrices along the loop is
\begin{align}\label{dm1}
\rho(t)\equiv\rho_{\theta(t)}=\sum_{m}\lambda_m| \psi_m(t)\rangle\langle \psi_m(t)|,
\end{align}
where $\lambda_m=\me^{-\beta m\omega_0}/Z$, and $| \psi_m(t)\rangle\equiv | \psi_m(\theta(t),0)\rangle$. The corresponding amplitude is
\begin{align}\label{dm2}
W(t)\equiv W_{\theta(t)}=\sum_{m}\sqrt{\lambda_m}| \psi_m(t)\rangle\langle \psi_m(t)|U(t).
\end{align}
Here $U(t)=\me^{-\mi\chi\mathlarger{\int}_0^t\theta'\dif t' J_y}$ with $\theta'=\frac{\dif \theta(t')}{\dif t'}$.
If $t$ denotes the time as the Uhlmann process is simulated by a time evolution process, $W(t)$ cannot satisfy the parallel condition (\ref{Wpt1}) due to the incompatibility between the dynamical and Uhlmann processes~\cite{ourPRB20}. A workaround is to introduce a suitable dynamical process to compensate for the effect of $t$. Here we include an extra time evolution governed by the Hamiltonian $H_\text{S}=J_y\dot{\theta}$ with the time-evolution operator
\begin{align}\label{US}U_\text{S}(t)=\me^{-\mi\mathlarger{\int}_0^tH_\text{S} (t')\dif t' }=\me^{-\mi\mathlarger{\int}_0^t\theta'\dif t' J_y},\end{align} inspired by Ref.~\cite{npj18}. Thus, the amplitude takes the form
\begin{align}\label{dm2b}
W(t)=\sum_{m}\sqrt{\lambda_m}U_\text{S}(t)| \psi_m(t)\rangle\langle \psi_m(t)|U^\dagger_\text{S}(t)U(t).
\end{align}

Experimentally, one may simulate the Uhlmann process of a spin-$j$ system by the purified state associated with the amplitude as follows.
\begin{align}\label{dm3}
|W(t)\rangle=\sum_{m}\sqrt{\lambda_m}U_\text{S}(t)|\psi_m(t)\rangle\otimes U_\text{A}(t) |\psi_m(t)\rangle,
\end{align}
where the time-evolution operator of the ancilla is
\begin{align}\label{dm3b}
U_\text{A}(t)=\left[U^\dagger_\text{S}(t)U(t)\right]^T=\me^{-\mi\eta\mathlarger{\int}_0^t\theta'\dif t' J_y}.
\end{align}
Here $\eta=1-\chi=\text{sech}(\beta\omega_0/2)$ and $J^T_y=-J_y$ has been applied.
The purified state may be thought of as one living in an enlarged Hilbert space $\mathcal{H}=\mathcal{H}_\text{S}\otimes\mathcal{H}_\text{A}$ with S and A standing for the system and ancilla. $U_\text{S}(t)$ and $U_\text{A}(t)$ are the effective time evolution operators in $\mathcal{H}_\text{S}$ and $\mathcal{H}_\text{A}$, respectively. Correspondingly, $\eta$ can be recognized as the ancilla weight~\cite{npj18}.
The density matrix can be obtained by taking the partial trace over the ancilla: $\rho(t)=\text{Tr}_\text{A}\left(|W(t)\rangle\langle W(t)|\right)$. We remark that the need for a reservoir to keep the density matrix of the system the same in an Uhlmann process has been achieved by the entangled state~\eqref{dm3} encoding the temperature effects in its initial state, so the entangled system no longer needs a reservoir and follows the dynamics governed by $U_{S,A}$.
Experimentally, the operators $U_\text{S, A}$ may be realized in atomic simulators by applying radio-frequency pulses to the system and ancilla states to induce rotations along the $y$-axis that mimic the corresponding time-evolution.

Next, we verify that $|W(t)\rangle$ indeed satisfies the parallel-transport condition (\ref{dm0a}). A straightforward evaluation shows
\begin{widetext}
 \begin{align}\label{dm0d}
\text{Im}\langle W(t)|\frac{\dif}{\dif t}|W(t)\rangle&=\text{Im}\sum_{m,n=-j}^j\sqrt{\lambda_m\lambda_n}\Big(\langle\psi_n|U_\text{S}^\dagger\dot{U}_\text{S}|\psi_m\rangle\langle\psi_m|U_\text{A}U_\text{A}^\dagger|\psi_n\rangle+\langle\psi_n|U_\text{S}^\dagger U_\text{S}|\psi_m\rangle\langle\psi_m|\dot{U}_\text{A}U_\text{A}^\dagger|\psi_n\rangle\notag\\
&\qquad\qquad\qquad\qquad\quad+\langle\psi_n|U_\text{S}^\dagger U_\text{S}|\dot{\psi}_m\rangle\langle\psi_m|U_\text{A}U_\text{A}^\dagger|\psi_n\rangle+\langle\psi_n|U_\text{S}^\dagger U_\text{S}|\psi_m\rangle\langle\dot{\psi}_m|U_\text{A}U_\text{A}^\dagger|\psi_n\rangle\Big)\notag\\
&=-\text{Im}\sum_{m=-j}^j\lambda_m\mi\chi\dot{\theta}\langle\psi_m|J_y|\psi_m\rangle, \notag\\
&=0.
\end{align}
\end{widetext}
 where we have applied
  \begin{align}\label{dm0c}
\dot{U}_\text{S}=-\mi\dot{\theta}J_yU_\text{S},\quad \dot{U}_\text{A}=-\mi\eta\dot{\theta}J_yU_\text{A}
\end{align}
and $[J_y,U_\text{S}]=[J_y,U_\text{A}]=0$. Here $|\psi_m\rangle\equiv |\psi_m(\theta,0)\rangle=\me^{-\mi\theta J_y}|jm\rangle$.
Therefore, $|W(t)\rangle$ is indeed parallel transported along the meridian. The evaluation of Eq.~(\ref{dm0d}) involves some subtleties of the inner product in $\mathcal{H}_\text{A}$, which are summarized in Appendix~\ref{appc}.

When $t$ reaches $1$, the process completes a cycle as the amplitude is parallel transported according to Eq.~(\ref{dm0d}). However, to confirm that an Uhlmann process has been simulated, we need to show that the overlap between the initial and final purified states reproduces the Uhlmann fidelity.
Given $U_\text{S}(1)=\me^{-\mi2\pi\Omega J_y}$ and $U_\text{A}(1)=\me^{-\mi2\eta\pi\Omega J_y}$, we have
\begin{align}\label{GU1}
&\mathcal{G}^U(T,1)=\langle W(0)|W(1)\rangle\notag\\
&=\sum_{m,n=-j}^j\sqrt{\lambda_m\lambda_n}\langle jm|U_\text{S}(1)|\psi_n(1)\rangle\langle\psi_n (1)|U_\text{A}^T(1)|jm\rangle.
\end{align}
By applying $|\psi_n(1)\rangle=\me^{-\mi2\pi J_y}|jn\rangle$, it can be shown that  \begin{align}\langle jm|U_\text{S}(1)|\psi_n(1)\rangle=d^j_{mn}(2\pi(\Omega+1))=(-1)^{2j(\Omega+1)}\delta_{mn},\end{align}
which implies that only the $m=n$ terms give non-zero contributions to the sum. Thus, we can replace $\lambda_n$ by $\lambda_m$ in the second line of Eq.~(\ref{GU1}) and get
\begin{align}\label{GU2}
\mathcal{G}^U(T,1)
&=\sum_{m=-j}^{j}\frac{ \mathrm{e}^{-\beta \omega_{0} m}}{Z(0)}d^j_{mm}(2\pi\Omega\chi).
\end{align}
The result agrees with Eq.~(\ref{G1}) or (\ref{G2}). Therefore, the Uhlmann process of a spin-$j$ system may be simulated by using the purified states with the help of the ancilla. Finally, the argument of the Uhlmann fidelity $\theta_U=\arg\left[ \mathcal{G}^U(T,1)\right]$ gives the Uhlmann phase.

\subsection{Measuring TQPT using atomic simulator}
In the augmented system with the ancilla, the Uhlmann phase is  the relative phase between the initial and final purified states. A jump of the Uhlmann phase indicates the occurrence of a TQPT. Experimentally, the initial purified state of Eq.~\eqref{dm3} may be prepared by entangling two atoms, one as the system qu$j$it and the other as the ancilla qu$j$it. The coefficients $\lambda_m$ determines the temperature of the system in the mixed state. Then, the two atoms evolve according to the time evolution operators $U_S$ and $U_A$ that may be engineered from the Hamiltonians of the system and ancilla, respectively. After the time evolution produces a relative phase of the composite system equivalent to the Uhlmann phase, an interferometry of the evolved purified state $|W(1)\rangle$ with another identically prepared initial purified state $|W(0)\rangle$ may reveal if a nontrivial value of the Uhlmann phase has been accumulated. When the overlap between the initial and final purifications changes signs due to the Uhlmann phase acquired during the Uhlmann process, it indicates a TQPT as the Loschmidt amplitude vanishes. Since the initial entangled state $|W(0)\rangle$ already encodes the temperature effect, there is no need to introduce an external reservoir to the system plus ancilla because the time-evolution operators $U_S$ and $U_A$ are sufficient to generate the Uhlmann phase as if the system has been kept in equilibrium.

It is also possible to  experimentally investigate the geometrical generating function $g$ from the Loschmidt amplitude/ Uhlmann fidelity shown in Eq.~(\ref{g}). Recently, the DQPTs induced by a quantum quench have been experimentally studied by observing the Fisher zeros (related to dynamical vortices) of the Loschmidt amplitude~\cite{DQPTB4} or by measuring the non-analytic behaviors of the rate function, which is the counterpart of the dynamical free energy~\cite{DQPTB41}. The TQPTs in the Uhlmann processes also correspond to the zeros of the Loschmidt amplitude, analogous to the DQPTs after a quench. We expect those experimental techniques may be applicable to the experimental investigations of the TQPTs in the Uhlmann process. We briefly discuss the second method here. In Ref.~\cite{DQPTB41}, the rate function is defined as
\begin{align}
\gamma(t)=-\frac{1}{L}\ln|\mathcal{G}(t)|,
\end{align}
where $t$ is the time, and $\mathcal{G}(t)$ is the Loschmidt amplitude in a quench process. In a real experiment, $\mathcal{G}(t)$ is replaced by the probability of
returning to the initial manifold after a duration of $t$. For the spin-$j$ system simulated by atomic states discussed above, the system and acilla are formed by two atoms. The control parameter of the Uhlmann process is the temperature $T$. After evolving the system plus ancilla, the Loschmidt amplitude $\mathcal{G}^U(T)$ may be inferred from the probability of
returning to the initial purified state of the qu$j$its at temperature $T$ in an Uhlmann process.

\subsection{Measuring TQPT using digital quantum simulation}
Finally, we present an alternative experimental procedure for measuring the TQPTs of general spin-$j$ systems via digital quantum simulations by using standard qubits, or two-level systems. Previously, we mentioned that qu$j$its may be realized by the hyperfine states of atoms. Here we consider a different scheme for simulating the mixed states of spin-$j$ systems by purified states constructed from qubits. In principle, a quantum computer may be used to simulate the spin-$j$ system and reveal its TQPTs
by the following protocol, which is different from the protocol of Ref.~\cite{npj18} because the latter does not have a straightforward generalization to $n$-level systems.

The first step is to initialize a given purified state of the spin-$j$ system in a quantum register made of qubits. Suppose an integer $n$ satisfies $2^{n-1}\le 2j+1<2^n$,  then $2n$ is the minimal number of qubits to store an initial purified state
\begin{align}
|W(0)\rangle=\sum_m\sqrt{\lambda_m}|\psi_m(0)\rangle\otimes |\psi_m(0)\rangle.
\end{align}
Here the first state on the right hand side is the system and the second is the ancilla, and $\lambda_m=\me^{-\beta m\omega_0}/Z$ depends on temperature. Note that $U_\text{S}(0)=U_\text{A}(0)=1$ in Eq.~(\ref{dm3}).
Explicitly, we encode the states of the spin components as follows.
\begin{align}
|\psi_{-j}(0)&\rangle\rightarrow |00\cdots0\rangle, \notag\\
|\psi_{-j+1}(0)&\rangle\rightarrow |00\cdots1\rangle, \notag\\
&\vdots\notag\\
|\psi_{j}(0)&\rangle\rightarrow |j_1j_2\cdots j_n\rangle,
\end{align}
where $j_1 j_2 \cdots j_n$ is the binary representation of the number $2j+1$ with $j_1,j_2,\cdots, j_n=0,1$. In the qubit system, the initial purified state is mapped to
\begin{align}\label{W0}
|W(0)\rangle=\sum_{i=0}^{2^n-1}\sqrt{p_i}|i\rangle\otimes |i\rangle
\end{align}
where $p_0=\lambda_{-j}$, $p_1=\lambda_{-j+1}$, $\cdots$, $p_{2j+1}=\lambda_j$, and $p_{2j+2}=\cdots=p_{2^n-1}=0$.
The first (second) $n$ qubits carry a representation of the system (ancilla) state.
However, since we are using purified states of the system plus ancilla to simulate the behavior of the mixed states of the system, temperature determines the initial purified state but does not further enter into the experimental manipulations. In other words, $\lambda_m$ stands for a set of preassigned parameters, which characterize the mixed state of the system. With fixed values of $\lambda_m$, the system is effectively in equilibrium with temperature $T$.

Next, we need to prepare the state~(\ref{W0}) by qubits. This can be achieved by following the steps shown in Ref.~\cite{GLLPRA01}, which provides an efficient scheme to initialize an arbitrary superposed state in a quantum register. The scheme only involves the one-bit rotation and the controlled$^k$ gate, which has $k$ control qubits. The protocol is summarized in Appendix~\ref{app:d}. Once the input state is initialized, one may follow the time evolution and measure the outcome. We outline the procedure as follows.

\textit{Step 1}. Prepare the initial state $|W(0)\rangle$ by using qubits based on the above discussion and Appendix \ref{app:d}. For simplicity, we choose  $\theta(0)=0$ in the parameter space.

\textit{Step 2}. Consider a time evolution as the  system moves along the meridian in the parameter space according to $\theta'(t)=\frac{\dif \theta(t)}{\dif t}=vt$. Apply the unitary evolution $U_\text{S}(t)\otimes U_\text{A}(t)$ on the system and ancillary states, where the evolution is determined by Eqs.~(\ref{US}) and (\ref{dm3b}). The purified state then evolves according to Eq.~(\ref{dm3}). Importantly, Eq.~(\ref{dm0d}) guarantees that the process follows parallel transport.

\textit{Step 3}. After a cyclic process is completed ($\theta(1)=2\pi\Omega$ in case of $v=2\pi\Omega$), either the Uhlmann phase $\theta_U$ or the geometrical generating function $g$ can be obtained from the Uhlmann fidelity $\langle W(0)|W(1)\rangle$. For example, the Uhlmann phase appears as the relative phase between the initial and final purified states, which may be determined by interferometric techniques.

\textit{Step 4}. Change the parameter $\lambda_m$, which tunes the temperature $T$, and prepare another initial state $|W(0)\rangle$. Repeat the above steps, and a curve of $\theta_U$ as a function of $T$ is obtained. A jump of the Uhlmann phase signals the occurrence of a TQPT.

\section{Conclusion}
The general expressions of the Loschmidt amplitude and Uhlmann phase of a spin-$j$ paramagnet influenced by a magnetic field in an Uhlmann process have demonstrated the usefulness of characterizing finite-temperature topological properties via the Uhlmann process. To obtain compact expressions for a deeper understanding, we consider the system in thermal equilibrium traversing a great circle in the parameter space, so the path-ordered integration can be carried out. By analyzing specific examples with $j=\frac{1}{2}$ and $1$, we visualize the TQPTs at finite temperatures, indicated by quantized jumps of the Uhlmann phase. The number of TQPTs in the Uhlmann process is associated with the winding number in the parameter space and reveals the topological properties via the Uhlmann holonomy. In addition to topological regimes extended from the zero-temperature point, we found finite-temperature topological regimes, where nontrivial value of the Uhlmann phase is only possible at finite temperatures but not at zero temperature. In contrast to the DQPT of a quench process dealing with non-equilibrium systems, there is no infinite-temperature TQPT in a Uhlmann process. With the rapid progress in quantum simulations and sensing, the framework of Uhlmann process will help advance our understanding of the interplay between topological properties and finite-temperature effects.

\acknowledgments{
H. G. was supported by the National Natural Science Foundation
of China (Grant No. 12074064). C. C. C. was supported by the National Science Foundation under Grant No. PHY-2011360.} X.-Y. Hou was supported by the Scientific
Research Foundation of the Graduate School of Southeast University (Grant No. YBPY2029).

\appendix

\section{Details of some derivations}\label{appa}

\subsection{Uhlmann connection of spin-$j$ system}
By using the The Campbell-Baker-Hausdorff formula $\me^{\hat{A}}\hat{B}\me^{-\hat{A}}=\hat{B}+[\hat{A},\hat{B}]+\frac{1}{2!}[\hat{A},[\hat{A},\hat{B}]]+\cdots$, the following results can be obtained
\begin{align}
\me^{-\mi\theta J_y}J_z\me^{\mi\theta J_y}&
=J_x\sin\theta+J_z\cos\theta, \label{het1}\\
\me^{-\mi\phi J_z}J_x\me^{\mi\phi J_z}&=J_x\cos\phi+J_y\sin\phi, \label{het2}\\
\me^{-\mi\phi J_z}J_y\me^{\mi\phi J_z}&=-J_x\sin\phi+J_y\cos\phi, \label{het2a}\\
\me^{-\mi\theta J_y}J_x\me^{\mi\theta J_y}&=-J_z\sin\theta+J_x\cos\theta.\label{het1a}
\end{align}
Thus, the first line of Eq.(\ref{sjH1}) can be proved as follows
\begin{align}\label{het3}
&\me^{-\mi\phi J_z}\me^{-\mi\theta J_y}J_z\me^{\mi\theta J_y}\me^{\mi\phi J_z}\notag\\
&=J_x\sin\theta\cos\phi+J_y\sin\theta\sin\phi+J_z\cos\theta.
\end{align}
The last line of Eq.(\ref{sjH1}) is quite straightforward since $[\me^{\mi\phi J_z},J_z]=0$.

To prove Eq.(\ref{srho}), applying the expression (\ref{sjH1}) of the Hamiltonian and $\rho=\frac{1}{Z}\me^{-\beta H}$, we get
\begin{align}\sqrt{\rho}=\frac{\me^{-\frac{\beta H}{2}}}{\sqrt{Z}}=\frac{1}{\sqrt{Z}}\me^{-\frac{1}{2}\beta V\omega_{0}J_{z}V^{\dagger}}=\frac{1}{\sqrt{Z}}V\me^{-\frac{1}{2}\beta \omega_{0}J_{z}}V^{\dagger}.\end{align} The differential of $\sqrt{\rho}$ is
\begin{align}
\dif\sqrt{\rho}&=\frac{\dif V\me^{-\frac{\beta \omega_{0}J_{z}}{2}}V^{\dagger}}{\sqrt{Z}}+\frac{V\me^{-\frac{\beta \omega_{0}J_{z}}{2}}\dif V^{\dagger}}{\sqrt{Z}}
-\frac{1}{2Z^{\frac{3}{2}}}\dif Z\me^{-\frac{\beta H}{2}} \notag \\
&=\frac{\dif VV^{\dagger}\me^{-\frac{\beta H}{2}}}{\sqrt{Z}}+\frac{\me^{-\frac{\beta H}{2}}V\dif V^{\dagger}}{\sqrt{Z}}-\frac{1}{2Z^{\frac{3}{2}}}\dif Z\me^{-\frac{\beta H}{2}}.
\end{align}
The last term commutes with $\sqrt{\rho}$, thus it is straightforward to show
\begin{align}\label{srho2}
[\dif\sqrt{\rho},\sqrt{\rho}]
=\frac{\{\dif VV^{\dagger},\me^{-\beta H}\}}{Z}+\frac{2\me^{-\frac{\beta H}{2}}V\dif V^{\dagger}\me^{-\frac{\beta H}{2}}}{Z} .
\end{align}

To prove Eq.~(\ref{UdU}), we note $V^\dagger=\me^{-\mi\phi J_z}\me^{\mi\theta J_y}\me^{\mi\phi J_z}$ and
\begin{align}\label{UdUp}
\dif V^\dagger &=\mi\me^{-\mi\phi J_z}\me^{\mi\theta J_y} J_y\me^{\mi\phi J_z}\dif\theta-\mi[J_z,V^\dagger]\dif \phi.
\end{align}
To evaluate the first term on the right-hand-side, we consider an arbitrary function $f(J_y)$ which can be expressed as the power series of $J_y$ as
\begin{align}\label{f0}
f(J_y)=a_0+a_1J_y+a_2J^2_y+\cdots.
\end{align}
By applying Eq.~(\ref{het2a}), we have
\begin{align}\label{f1}
&\me^{-\mi\phi J_z}f(J_y)\me^{\mi\phi J_z}\notag\\&=a_0+a_1\me^{-\mi\phi J_z}J_y\me^{\mi\phi J_z}+a_2(\me^{-\mi\phi J_z}J_y\me^{\mi\phi J_z})^2+\cdots\notag\\
&=f(-J_x\sin\phi+J_y\cos\phi).
\end{align}
By Eq.~(\ref{f1}), the first term on the right hand side of Eq.~(\ref{UdUp}) becomes \begin{align}&\mi\me^{\mi\theta(-J_x\sin\phi+J_y\cos\phi)}(-J_x\sin\phi+J_y\cos\phi)\dif\theta\notag\\&=-\mi V^\dagger(J_x\sin\phi-J_y\cos\phi)\dif\theta.\end{align} Hence, we get
\begin{align}\label{UdU2}
 V\dif V^{\dagger}=-\mi(J_{x}\sin\phi-J_{y}\cos\phi)\dif \theta+ \mi (J_{z}-VJ_zV^\dagger)\dif \phi.
\end{align}

\subsection{Uhlmann curvature and Chern number}\label{app:curvature}
Applying the relation $\mathbf{J}\times\mathbf{J}=\mi\mathbf{J}$ for the spin-$j$ system, the Uhlmann curvature becomes
\begin{align}\label{FU0}
F_U&=\mi(2\chi-\chi^2)[(J_x\cos\phi+J_y\sin\phi)\sin\theta\notag\\&+J_z\cos\theta]\sin\theta\dif\theta\wedge\dif\phi\notag\\
&=\frac{\mi(2\chi-\chi^2)}{\omega_0}H(\theta,\phi)\sin\theta\dif\theta\wedge\dif\phi.
\end{align}
Since $2\chi-\chi^2=1-(1-\chi)^2=\tanh^2\frac{\beta\omega_0}{2}$, the Uhlmann curvature has the expression 
\begin{align}\label{FU1}
F_U
&=\mi\frac{H(\theta,\phi)}{\omega_0}\tanh^2\left(\frac{\beta\omega_0}{2}\right)\sin\theta\dif\theta\wedge\dif\phi,
\end{align}
where $2\chi-\chi^2=1-(1-\chi)^2=\tanh^2\frac{\beta\omega_0}{2}$ has been applied.
Since $\text{Tr}J_x=\text{Tr}J_y=\text{Tr}J_z=0$,
the Chern number $\text{Ch}_U=\frac{\mi}{2\pi}\mathlarger{\int}\text{Tr}F_U =0$.

\subsection{Uhlmann process of a pure state}
When undergoing an Uhlmann process parameterized by $t$, the density matrix of a pure state is $\rho(t)=|\psi(t)\rangle\langle\psi(t)|$. However, $|\psi(t)\rangle$ can not always be normalized with $\langle\psi(t)|\psi(t)\rangle=1$ in an Uhlmann process because the condition $\text{Tr}\rho=1$ is not preserved by the Uhlmann process~\cite{ourPRB20}. If we let $\lambda=\langle\psi|\psi\rangle$, then $\rho|\psi\rangle=\lambda|\psi\rangle$. We normalize $|\psi\rangle$ as $|\tilde{\psi}\rangle=\frac{1}{\sqrt{\lambda}}|\psi\rangle$. Thus, \begin{align}
\rho=\lambda|\tilde{\psi}\rangle\langle\tilde{\psi}|,\quad \sqrt{\rho}=\sqrt{\lambda}|\tilde{\psi}\rangle\langle\tilde{\psi}|.\end{align}
Using $\langle\tilde{\psi}|\tilde{\psi}\rangle=1$, the commutator in Eq.~(\ref{GrAU}) is given by
\begin{align}\label{dr}
[\dif\sqrt{\rho},\sqrt{\rho}]&=\lambda\dif|\tilde{\psi}\rangle\langle\tilde{\psi}|+\lambda|\tilde{\psi}\rangle(\dif\langle\tilde{\psi}|)|\tilde{\psi}\rangle\langle\tilde{\psi}|\notag\\
&-\lambda|\tilde{\psi}\rangle\dif\langle\tilde{\psi}|-\lambda|\tilde{\psi}\rangle\langle\tilde{\psi}|(\dif|\tilde{\psi}\rangle)\langle\tilde{\psi}|.
\end{align}
Substitute this into Eq.~(\ref{GrAU}), we get
\begin{align}\label{AUj0}
A_U=0
\end{align}
for a pure state in an Uhlmann process.
According to Eq.~(\ref{Up}), we always have $\theta_U=0$.

\subsection{Inner product in the ancillary space}\label{appc}
In the main text, it is pointed out that the inner product between two purified states is isomorphic to the Hilbert-Schmidt product between two amplitudes:
\begin{align}\label{ipe1}
\langle W_1|W_2\rangle=\textrm{Tr}(W^\dagger_1W_2).
\end{align}
The verification of this identity involves a  proper treatment of the inner product in the ancillary space.
To see this, we consider a simple situation where $W_1$ and $W_2$ are two different amplitudes of the same density matrix $\rho$: $W_{1,2}=\sqrt{\rho}U_{1,2}$, where $\rho=\sum_n\lambda_n|n\rangle\langle n|$. Thus, it is straightforward to show that
\begin{align}\label{ipe2}
\textrm{Tr}(W^\dagger_1W_2)=\textrm{Tr}(U^\dagger_1\sqrt{\rho}\sqrt{\rho}U_2)=\textrm{Tr}(\rho U_2U^\dagger_1).
\end{align}
The corresponding purified states $|W_{1,2}\rangle$ are given by
\begin{align}\label{w2a}
|W_1\rangle&=\sum_n\sqrt{\lambda_n}|n\rangle\otimes U^T_1|n\rangle,\notag\\
|W_2\rangle&=\sum_m\sqrt{\lambda_m}|n\rangle\otimes U^T_2|m\rangle.
\end{align}
Here $U^{T}_{1,2}$ is the transpose of $U_{1,2}$, and $|n\rangle$ should be understood as the transpose of $\langle n|$ with no complex conjugation imposed. Otherwise, the left-hand-side of Eq.~(\ref{ipe1}) would be evaluated as follows.
\begin{align}\label{ipe3}
\langle W_1|W_2\rangle&=\sum_{n,m}\sqrt{\lambda_n\lambda_m}\langle n|m\rangle \langle n|U^\ast_1 U^T_2|m\rangle\notag\\
&=\sum_n \langle n|\lambda_nU^\ast_1 U^T_2|n\rangle\notag\\
&=\textrm{Tr}(\rho U^\ast_1 U^T_2),
\end{align}
where $\langle n|m\rangle =\delta_{mn}$ has been used.
This contradicts Eq.~(\ref{ipe3}). In fact, the overlap under the transposition operation must be evaluated as follows.
\begin{align}\label{ipe4}
\langle W_1|W_2\rangle&=\sum_{n,m}\sqrt{\lambda_n\lambda_m}\langle n|m\rangle \langle m|U_2 U^\dagger_1|n\rangle\notag\\
&=\sum_n \lambda_n\langle n|\sum_m|m\rangle\langle m|U_2 U^\dagger_1|n\rangle\notag\\
&=\textrm{Tr}(W^\dagger_1W_2).
\end{align}
In the main text, Eqs.~(\ref{dm0d}) and (\ref{GU1}) are both evaluated in this manner.

\section{Preparation of arbitrary initial state}\label{app:d}
Here we present a protocol for initializing the state (\ref{W0}) by $2n$ qubits based on the key idea of Ref.~\cite{GLLPRA01}. To simplify the notation, we abbreviate $|W(0)\rangle$ as
\begin{align}\label{W0a}
|W(0)\rangle=\sum_{i=0}^{2^n-1}\sqrt{p_i}|i\rangle\rangle,
\end{align}
where $|i\rangle\rangle\equiv |i\rangle\otimes |i\rangle$. Since $\sum_{i=0}^{2^n-1}p_i=1$, there are $2^n-1$ independent weights among $p_i$'s. We parameterize the $p_i$'s as
\begin{align}\label{pi}
\sqrt{p_0}&=\cos\alpha_1\cos\alpha_2\cdots\cos\alpha_{2^n-2}\cos\alpha_{2^n-1},\notag\\
\sqrt{p_1}&=\cos\alpha_1\sin\alpha_2\cdots\cos\alpha_{2^n-2}\cos\alpha_{2^n-1},\notag\\
&\cdots\notag\\
\sqrt{p_{2^n-2}}&=\sin\alpha_1\sin\alpha_2\cdots\sin\alpha_{2^n-2}\cos\alpha_{2^n-1},\notag\\
\sqrt{p_{2^n-1}}&=\sin\alpha_1\sin\alpha_2\cdots\sin\alpha_{2^n-2}\sin\alpha_{2^n-1}.
\end{align}
Some $p_i$'s may be zero if $2j+1 <2^n$.
$|W(0)\rangle$ can be constructed from the state $|0\rangle\rangle\otimes\cdots\otimes|0\rangle\rangle=|0\cdots 0\rangle\otimes|0\cdots 0\rangle$ by single-bit rotations and $k$-bit controlled rotations, which will be illustrated by explicit examples.

\begin{figure}[th]
\centering
\includegraphics[width=1.7in]{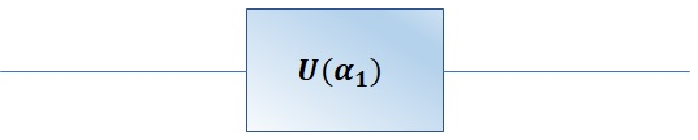}
\includegraphics[width=2.7in]{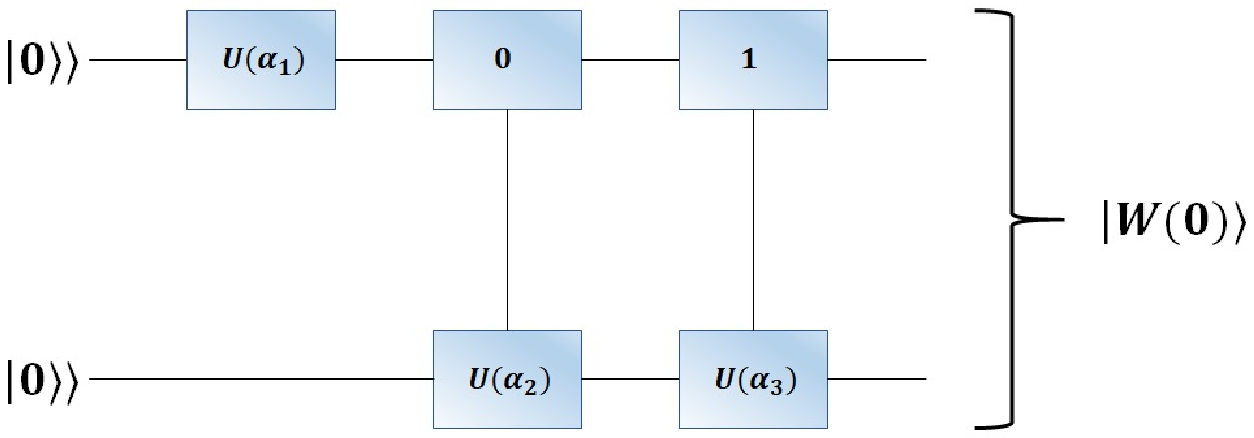}
\caption{(Top) The rotation operator  $U(\alpha_1)$. (Bottom) Quantum circuit for $n=2$.}
\label{QC2}
\end{figure}

If $n=1$, only one rotation is needed:
\begin{align}\label{n1}
U(\alpha_1)|0\rangle\rangle=\cos\alpha_1|0\rangle\rangle+\sin\alpha_1|1\rangle\rangle=|W(0)\rangle,
\end{align}
which is shown in the top panel of Fig.~\ref{QC2}. Here $U(\alpha_1)$ actually acts on the product space of the system and ancilla. For convenience, we still refer to it as a single-bit rotation.

\begin{figure}[th]
\centering
\includegraphics[width=3.8in]{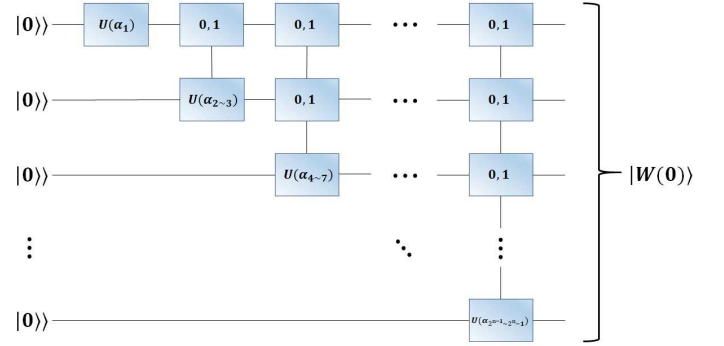}
\caption{Quantum circuit for preparing an state with $n$ qubits.}
\label{QC3}
\end{figure}

If $n=2$, we have
\begin{align}\label{n2}
|W(0)\rangle&=\cos\alpha_1\cos\alpha_2|00\rangle\rangle+\cos\alpha_1\sin\alpha_2|01\rangle\rangle\notag\\
&+\sin\alpha_1\cos\alpha_3|10\rangle\rangle+\sin\alpha_1\sin\alpha_3|11\rangle\rangle\notag\\
&=\cos\alpha_1|0\rangle\rangle\left(\cos\alpha_2|0\rangle\rangle+\sin\alpha_2|1\rangle\rangle\right)\notag\\
&+\sin\alpha_1|1\rangle\rangle\left(\cos\alpha_3|0\rangle\rangle+\sin\alpha_3|1\rangle\rangle\right).
\end{align}
Thus, $|W(0)\rangle$ can be realized by the quantum circuit shown in the bottom of Fig.~\ref{QC2}, where one single-bit rotation $U(\alpha_1)$ and two controlled gate operations are needed. To achieve this, two qubits are initialized in the state $|0\rangle\rangle\otimes|0\rangle\rangle$. We then perform $U(\alpha_1)$ to the first qubit,
which also acts as the control qubit. The second qubit is the target qubit. If the control qubit is in state $|0\rangle$ or $|1\rangle$ (labelled as 0 or 1 in the box), we perform $U(\alpha_{2})$ or $U(\alpha_{3})$ to the target qubit, respectively. Finally, the procedure applies to the generic case of $n$ qubits with the quantum circuit visualized in Fig.~\ref{QC3}. We have used compact boxes with labels $0,1$ and $U(\alpha_{i-j})$ to represent the controlled rotations of Fig.~\ref{QC2}.

\bibliographystyle{apsrev}

\end{document}